\documentclass[lettersize,journal]{IEEEtran}
\usepackage{amsmath,amsfonts}
\usepackage{algorithmic}
\usepackage{algorithm}
\usepackage{array}
\usepackage[caption=false,font=normalsize,labelfont=sf,textfont=sf]{subfig}
\usepackage{textcomp}
\usepackage{stfloats}
\usepackage{url}
\usepackage{verbatim}
\usepackage{graphicx}
\usepackage{bm}
\usepackage{multirow}
\usepackage{graphicx}
\usepackage{booktabs}
\usepackage{threeparttable}
\usepackage{xcolor}

\usepackage{cite}
\newcommand{\HQC}[1]{\textsf{HQC-#1}}

\usepackage{longtable}

\usepackage{multirow}
\usepackage{tabularx}
\newcolumntype{L}[1]{>{\raggedright\arraybackslash}p{#1}}
\newcolumntype{C}[1]{>{\centering\arraybackslash}p{#1}}
\newcolumntype{R}[1]{>{\raggedleft\arraybackslash}p{#1}}

\usepackage{url}

\begin{document}

\title{Accelerating HQC for Post-Quantum TLS 1.3 \\ on x86 IoT Gateways}
\author{
Jihoon Jang, Hyunju Park, Jebin Kim, Seokhie Hong and Suhri Kim%
\thanks{Jihoon Jang, Hyunju Park, and Jebin Kim are with the
School of Cyber Security, Korea University, Seoul, Republic of Korea.\ \par
Suhri Kim is with the School of Mathematics, Statistics and
Data Science, Sungshin Women's University, Seoul, Republic of Korea (e-mail: suhrikim@sungshin.ac.kr).\ \par
Seokhie Hong is with SmartM2M, Republic of Korea.\ \par
Corresponding author: Suhri Kim (e-mail: suhrikim@sungshin.ac.kr).\ \par
This work has been submitted to the IEEE for possible publication. Copyright may be transferred without notice, after which this version may no longer be accessible.}%
}

\markboth{}%
{}

\maketitle

\begin{abstract}
Post-quantum TLS at an IoT gateway must protect many device connections without large increases in handshake delay, CPU cost, or network traffic. HQC provides code-based diversity beyond ML-KEM, but its computation and ciphertext sizes can increase these costs. We optimize HQC for x86 processors with AVX2, AVX-512, and the Galois Field New Instructions (GFNI), and integrate the resulting implementations into TLS 1.3. We extend branch-free Toom-Cook/Karatsuba multiplication to AVX-512, accelerate Reed--Solomon decoding with GFNI, improve Reed--Muller decoding, accelerate SHA3-512 and fixed-weight sampling, and port Frobenius additive FFT (FAFFT) multiplication to AVX-512 + GFNI for \HQC{5}. On an Intel Core i5-1135G7 processor, our AVX2 implementation reduces decapsulation by 11.1--21.5\% over the fastest prior AVX2 results. Our AVX-512 implementation reduces key generation by 24.9--29.5\%, encapsulation by 10.4--11.1\%, and decapsulation by 20.6--26.1\% relative to Cabral et al. across the three HQC parameter sets. We evaluate the effects of these implementations on TLS 1.3 handshake latency and server and client CPU costs, as well as the effects of round-trip time (RTT) and bandwidth. In local loopback TLS measurements, our AVX-512 \HQC{5} implementation reduces handshake latency from 3.22~ms to 2.92~ms compared with Cabral et al. On a constrained path with 1~Mbit/s bandwidth and an added RTT of 50~ms, the \HQC{5} handshake takes 243~ms, compared with 87~ms for ML-KEM-1024. This indicates that HQC public-key and ciphertext sizes, rather than implementation speed, determine the remaining handshake cost on constrained links.
\end{abstract}


\section{Introduction}
\IEEEPARstart {L}{arge-scale} quantum computers would break the classical public-key cryptography used by current secure communication protocols through Shor's algorithm~\cite{shor1999polynomial}. This threat is particularly important in IoT environments, where resource-constrained devices connect to external services through gateways. Deploying post-quantum cryptography in such environments requires efficient implementations that minimize the additional computational cost and communication latency introduced by the transition.

TLS is a representative deployment target for post-quantum key establishment. In TLS 1.3, a KEM determines the key-share group. Its public key and ciphertext are carried in the handshake flights. Thus, post-quantum TLS affects gateway computation and the latency of device connections. NIST standardized ML-KEM as FIPS~203 in 2024~\cite{nistFIPS203} and selected HQC~\cite{NISTPQC-R4:HQC22}, a code-based KEM, for standardization in March 2025 to provide diversity beyond lattice-based ML-KEM. The main HQC components are polynomial multiplication over $\mathbb{F}_2[x]/(x^N-1)$, Reed-Muller and Reed-Solomon decoding, and Keccak-based functions. Efficient implementations of these components are therefore important for practical post-quantum TLS at an IoT gateway.

AVX-512 has been applied to a wide range of cryptographic schemes~\cite{Alter2021SPHINCSAVX512,TCHES:CFGRR21,TCHES:CFGR22,awslabsBikeKem,Zheng2024MLKEMTLSAVX512,Xu2024DilithiumAVX512,EPRINT:ZhaZha25,Lee2026AIMerAVX512}. It extends x86 SIMD processing to 512-bit registers and provides a richer vector instruction set. VPCLMULQDQ supports vectorized carry-less multiplication, while GFNI supports SIMD arithmetic over $\mathbb{F}_{2^8}$. Recent processors such as Tiger Lake support AVX-512, VPCLMULQDQ, and GFNI.

Cabral et al.~\cite{cabral2026accelerating} presented an AVX-512 implementation of HQC with optimizations for polynomial multiplication, Keccak-based functions, and both Reed-Muller and Reed-Solomon decoding. However, their Reed-Solomon decoder does not use GFNI. It relies on PCLMULQDQ-based field multiplication, with SIMD parallelism applied to the syndrome and error-value computations while the remaining arithmetic is largely scalar. In addition, the SHA3-512 oracles \texttt{hash\_g} and \texttt{hash\_i} remain scalar, the AVX-512 SHAKE-256 expansion is not applied to key generation, and single polynomial multiplication and support expansion are still based on AVX2 implementations.

\IEEEpubidadjcol

HQC optimization and PQ-TLS evaluation have largely been studied separately. To our knowledge, existing PQ-TLS studies compare HQC with other KEMs using the implementation provided by the TLS stack~\cite{Sosnowski23, Souvatzidaki25}. They do not vary the HQC implementation to determine how its optimization changes protocol cost. We optimize HQC for AVX2 and AVX-512 + GFNI, integrate the resulting implementations into one fixed TLS 1.3 stack, and measure their effects on handshake latency and server and client CPU costs. The network experiments then show when communication costs imposed by round-trip time (RTT) and bandwidth dominate these computation gains.

\subsection{Contributions}

We optimize polynomial multiplication, decoding, Keccak-based functions, and fixed-weight sampling using AVX-512 and GFNI. We also apply our decoding methods to AVX2. The main contributions of this paper are summarized as follows.

\begin{itemize}

\item \textbf{Polynomial multiplication on AVX-512.} We extend the branch-free Toom-Cook/Karatsuba design of Jang et al.~\cite{Jang26} to AVX-512 and select an efficient decomposition tree for each HQC parameter set. The resulting implementation has cycle counts within about 1\% of Robert and V\'eron~\cite{RV23} at \HQC{1} and \HQC{3}, and reduces the cost by 14.1\% at \HQC{5}. We also port the FAFFT multiplication of Chen et al.~\cite{cryptoeprint:2026/014} to AVX-512 + GFNI and reduce its transform cost with precomputed twiddle tables and reduced basis conversions. At \HQC{5}, the optimized FAFFT implementation reduces the cycle count by 17.7\% over Robert and V\'eron.

\item \textbf{Reed-Solomon and Reed-Muller decoding.} 
For Reed-Solomon decoding, we implement GFNI-based arithmetic for both AVX2 and AVX-512 and apply it to the critical arithmetic loops in all five $\mathbb{F}_{2^8}^{\mathrm{RS}}$ phases. Our AVX2 implementation reduces the decoder cycle count by 83.2\%--87.1\% over Chen et al.~\cite{cryptoeprint:2026/014}, while our AVX-512 implementation reduces it by 88.9\%--91.0\% over Cabral et al.~\cite{cabral2026accelerating}. For Reed-Muller decoding, we replace the binary search for the maximum in \texttt{find\_peaks} with a horizontal vector reduction in both AVX2 and AVX-512. This reduces the full decoder cost by 18.2\%--20.4\% over Chen et al. on AVX2 and by 3.7\%--9.9\% over Cabral et al. on AVX-512.

\item \textbf{Keccak-based functions and fixed-weight sampling.} We extend AVX-512 processing to operations that remain scalar or AVX2-based in the existing AVX-512 implementation. We implement the SHA3-512 oracles \texttt{hash\_g} and \texttt{hash\_i} with AVX-512, reducing their cycle counts by 37.4\%--46.4\% and 46.2\%--46.7\%, respectively, over Cabral et al.~\cite{cabral2026accelerating}. We also port \texttt{vect\_write\_support\_to\_vector}, which expands the sampled support into a dense vector, from AVX2 to AVX-512, reducing the support-expansion cost by 34.5\%--36.8\%.

\item \textbf{TLS 1.3 integration and assessment.} We integrate optimized HQC implementations into TLS 1.3 through liboqs and oqs-provider. Using identical protocol stacks, we measure their effects on handshake latency and server and client CPU costs, while distinguishing implementation-level improvements from the impact of network conditions such as RTT and bandwidth. The measurements identify the \HQC{5} ciphertext as the cause of an additional round trip at the default TCP initial congestion window.

\end{itemize}

At the KEM level, our AVX2 implementation reduces decapsulation by 11.1--21.5\% over the fastest prior AVX2 results of Jang et al.~\cite{Jang26} and Chen et al.~\cite{cryptoeprint:2026/014}. Relative to Cabral et al.~\cite{cabral2026accelerating}, our AVX-512 implementation reduces key generation by 24.9--29.5\%, encapsulation by 10.4--11.1\%, and decapsulation by 20.6--26.1\%. At \HQC{5}, the FAFFT configuration increases these reductions to 33.1\%, 16.3\%, and 27.4\%, respectively.

The source code and implementation details of the proposed method are available at \url{https://github.com/jihoonjang98/hqc-x86-simd}.

\subsection{Related Works}

The HQC NIST submission~\cite{NISTPQC-R4:HQC22} provides an optimized AVX2 implementation. It uses PCLMULQDQ for carry-less polynomial multiplication and for $\mathbb{F}_{2^8}^{\mathrm{RS}}$ multiplication in the Reed-Solomon decoder. In the latter, \texttt{gf\_mul\_vect} evaluates 16 products in parallel for syndrome computation, whereas several other arithmetic functions call \texttt{gf\_mul}, which computes one field multiplication using PCLMULQDQ. We use this implementation as the AVX2 baseline.

Robert and V\'eron~\cite{RV23} studied AVX-512 polynomial multiplication over $\mathrm{GF}(2)[x]$ using VPCLMULQDQ. Their implementation selects Toom-Cook/Karatsuba combinations according to the target degree range and applies them to the HQC parameter sets. Their work focuses on polynomial multiplication.

Jang et al.~\cite{Jang26} proposed GFmulOpt, a systematic search method for branch-free Toom-Cook/Karatsuba decomposition trees for $\mathrm{GF}(2)[x]$ polynomial multiplication. They evaluate GFmulOpt on AVX2 with PCLMULQDQ and VPCLMULQDQ, and on Arm Neon with PMULL. For \HQC{1} and \HQC{3} on AVX2, it reduces the polynomial multiplication cost by 17--32\% over the HQC baseline, and it shows that an additive-FFT approach may be preferable for \HQC{5}. Their work focuses on polynomial multiplication and does not address AVX-512.

Chen et al.~\cite{cryptoeprint:2026/014} applied Frobenius additive FFT (FAFFT) to HQC polynomial multiplication on x86 AVX2 and GFNI platforms. They also study a CRT-based additive-FFT decomposition for the non-power-of-two polynomial lengths in HQC, which is used in their implementation for platforms without long binary-polynomial multiplication instructions. For x86, they reuse transforms of common operands and cache transformed key polynomials to reduce repeated transform costs. They also optimize Reed-Muller and Reed-Solomon encoding and decoding on AVX2. The Reed-Solomon part further uses GFNI-based $\mathbb{F}_{2^8}^{\mathrm{RS}}$ arithmetic. The cached variant improves repeated encapsulation and decapsulation at the cost of increased local key storage.

Cabral et al.~\cite{cabral2026accelerating} proposed an AVX-512 implementation of HQC. Their implementation accelerates polynomial multiplication, Reed-Muller decoding, Reed-Solomon decoding, and Keccak-based functions. For polynomial multiplication, the single-product path retains the Toom-Cook/Karatsuba implementation of HQC at AVX2 width. For the two products sharing $\mathbf{r}_2$, they pack $\mathbf{h}$ and $\mathbf{s}$ into ZMM registers and process both products with AVX-512. However, the Reed-Solomon decoder does not use GFNI, and several $\mathbb{F}_{2^8}^{\mathrm{RS}}$ arithmetic functions remain scalar. Moreover, the SHA3-512 oracles and support expansion are not covered by their AVX-512 implementation.

Chen et al.~\cite{cryptoeprint:2026/1588} subsequently proposed the Strided Frobenius Additive FFT. SFAFFT introduces a stride to reduce the number of butterfly stages and performs the leaf products over smaller quotient rings. Their implementations target AVX2, GFNI, Apple M1, and Arm processors, but do not include AVX-512. The authors state that the source code will be released. Since it is currently unavailable, we leave a comparison under the same experimental conditions for future work.

Several works integrate post-quantum KEMs into TLS and measure handshake performance. Paquin et al.~\cite{PaqSteTam20} emulate network conditions and show that communication sizes determine handshake time on constrained links. Sosnowski et al.~\cite{Sosnowski23} measure 23 key-agreement variants, including HQC, on identical server hardware. They report additional round trips when server messages exceed the initial TCP congestion window. Tasopoulos et al.~\cite{Tasopoulos22} evaluate post-quantum TLS on embedded platforms. Souvatzidaki and Limniotis~\cite{Souvatzidaki25} evaluate HQC, Kyber, and BIKE in TLS 1.3 with the OQS provider under packet loss. These studies compare KEM choices using the implementations available in their respective TLS stacks. Zheng et al.~\cite{Zheng2024MLKEMTLSAVX512} optimize ML-KEM with AVX-512 and integrate it into TLS 1.3. Their ML-KEM operations are 1.64 times faster than AVX2, but their TLS handshake throughput improves by 3\%--5\%. In Section~\ref{sec:tls}, we compare optimized HQC implementations in one fixed TLS stack to measure their effect on protocol performance.

\subsection{Organization}

The remainder of this paper is organized as follows. Section~\ref{sec:preliminaries} reviews the HQC KEM, the AVX-512 + GFNI instruction set, and TLS 1.3 key exchange. Sections~\ref{sec:multiplication}--\ref{sec:keccak} describe the optimization targets in HQC and our corresponding methods. Section~\ref{sec:evaluation} presents the experimental results. Section~\ref{sec:tls} describes the TLS 1.3 integration and assesses protocol-level performance, and Section~\ref{sec:conclusion} concludes the paper.

\section{Preliminaries}
\label{sec:preliminaries}

This section introduces the notation used in the paper, reviews the HQC KEM, and summarizes the AVX-512 and GFNI instructions used throughout this work.

\subsection{Notation}
\label{ssec:notation}

Throughout the paper, $\mathbb{F}_2$ denotes the binary field. The Reed-Solomon code in HQC is defined over $\mathbb{F}_{2^8}^{\mathrm{RS}}=\mathbb{F}_2[x]/(x^8+x^4+x^3+x^2+1)$, whereas the GFNI multiplication instruction operates over $\mathbb{F}_{2^8}^{\mathrm{AES}}=\mathbb{F}_2[x]/(x^8+x^4+x^3+x+1)$. Let $R=\mathbb{F}_2[x]/(x^N-1)$ and let $R_w=\{\mathbf{v}\in R\mid |\mathbf{v}|=w\}$. For a vector $\mathbf{v}$ and an integer $n'\leq N$, $\mathsf{Trun}(\mathbf{v},n')$ keeps the low $n'$ bits of $\mathbf{v}$.

Bold lowercase letters, such as $\mathbf{h}$ and $\mathbf{r}_2$, denote elements of $R$ or code vectors. The parameters $N$, $n_1$, $n_2$, $\omega$, $\omega_r$, $\omega_e$, and $\delta$ follow Table~\ref{tab:hqc-parameters}. The oracles $H$ and $J$ are instantiated with SHA3-256, while $G$ and $I$ are instantiated with SHA3-512. We denote the extendable-output function by $\mathtt{XOF}$.

\subsection{HQC}
\label{ssec:hqc}

HQC~\cite{NISTPQC-R4:HQC22} is a code-based KEM based on the hardness of syndrome decoding for quasi-cyclic codes. It is obtained by applying an implicit-rejection transform to an IND-CPA-secure public key encryption scheme, denoted by \textsf{HQC-PKE}. The three parameter sets, \HQC{1}, \HQC{3}, and \HQC{5}, target NIST security levels 1, 3, and 5, respectively.

\begin{table}[!h]
\centering
\caption{HQC parameter sets~\cite{NISTPQC-R4:HQC22}.}
\label{tab:hqc-parameters}
\begin{tabular}{lccccccc}
\toprule
Parameter set & Security & $N$ & $n_1$ & $n_2$ & $\delta$ & $\omega$ & $\omega_r,\omega_e$ \\
\midrule
\HQC{1} & NIST 1 & 17{,}669 & 46 & 384 & 15 & 66  & 75 \\
\HQC{3} & NIST 3 & 35{,}851 & 56 & 640 & 24 & 100 & 114 \\
\HQC{5} & NIST 5 & 57{,}637 & 90 & 640 & 29 & 131 & 149 \\
\bottomrule
\end{tabular}
\end{table}

HQC uses a concatenated code $\mathcal{C}$ of length $n_1 n_2$. The outer code is a Reed-Solomon code over $\mathbb{F}_{2^8}^{\mathrm{RS}}$, and the inner code is a repeated Reed-Muller code based on $\mathrm{RM}(1,7)$. The Reed-Solomon component has error-correction capacity $\delta$, while the Reed-Muller component repeats each encoded symbol $n_2/128$ times. Algorithms~\ref{alg:hqc-pke} and~\ref{alg:hqc-kem} summarize the components of HQC.

\begin{algorithm}[!h]
\caption{Main components in \textsf{HQC-PKE}~\cite{NISTPQC-R4:HQC22}.}
\label{alg:hqc-pke}
\linespread{1.2}\footnotesize
\begin{algorithmic}[1]
\STATE \textbf{Global parameters} $(N,n_1,n_2,\omega,\omega_r,\omega_e,\mathcal{C})$
\vspace{5pt}
\STATE \textsf{HQC-PKE.KeyGen}$(\mathit{seed}_{\mathsf{PKE}})$
\STATE \quad $(\mathit{seed}_{\mathsf{dk}},\mathit{seed}_{\mathsf{ek}}) \gets I(\mathit{seed}_{\mathsf{PKE}})$
\STATE \quad $(\mathbf{y},\mathbf{x}) \stackrel{\$}{\gets} R_\omega \times R_\omega$ from $\mathit{seed}_{\mathsf{dk}}$
\STATE \quad $\mathbf{h} \stackrel{\$}{\gets} R$ from $\mathit{seed}_{\mathsf{ek}}$
\STATE \quad $\mathbf{s} \gets \mathbf{x} + \mathbf{h}\cdot\mathbf{y}$
\STATE \quad \textbf{return} $\mathsf{ek}_{\mathsf{PKE}} \gets (\mathit{seed}_{\mathsf{ek}},\mathbf{s})$, $\mathsf{dk}_{\mathsf{PKE}} \gets \mathit{seed}_{\mathsf{dk}}$
\vspace{5pt}
\STATE \textsf{HQC-PKE.Encrypt}$(\mathsf{ek}_{\mathsf{PKE}},\mathbf{m},\theta)$
\STATE \quad $\mathbf{h} \stackrel{\$}{\gets} R$ from $\mathit{seed}_{\mathsf{ek}}$
\STATE \quad $(\mathbf{r}_2,\mathbf{e},\mathbf{r}_1) \stackrel{\$}{\gets} R_{\omega_r}\times R_{\omega_e}\times R_{\omega_r}$ from $\theta$
\STATE \quad $\mathbf{u} \gets \mathbf{r}_1 + \mathbf{h}\cdot\mathbf{r}_2$
\STATE \quad $\mathbf{v} \gets \mathcal{C}.\mathsf{Encode}(\mathbf{m}) + \mathsf{Trun}(\mathbf{s}\cdot\mathbf{r}_2+\mathbf{e}, n_1n_2)$
\STATE \quad \textbf{return} $\mathbf{c}_{\mathsf{PKE}} \gets (\mathbf{u},\mathbf{v})$
\vspace{5pt}
\STATE \textsf{HQC-PKE.Decrypt}$(\mathsf{dk}_{\mathsf{PKE}},\mathbf{c}_{\mathsf{PKE}}=(\mathbf{u},\mathbf{v}))$
\STATE \quad $\mathbf{y} \stackrel{\$}{\gets} R_\omega$ from $\mathit{seed}_{\mathsf{dk}}$
\STATE \quad \textbf{return} $\mathbf{m} \gets \mathcal{C}.\mathsf{Decode}(\mathbf{v}-\mathsf{Trun}(\mathbf{u}\cdot\mathbf{y}, n_1n_2))$
\end{algorithmic}
\end{algorithm}

\vspace{-0.5\baselineskip}

\begin{algorithm}[!h]
\caption{Main components in \textsf{HQC-KEM}~\cite{NISTPQC-R4:HQC22}.}
\label{alg:hqc-kem}
\linespread{1.2}\footnotesize
\begin{algorithmic}[1]
\STATE \textsf{HQC-KEM.KeyGen}$()$
\STATE \quad $\mathit{seed}_{\mathsf{KEM}} \stackrel{\$}{\gets} \{0,1\}^{|\mathit{seed}|}$
\STATE \quad $(\mathit{seed}_{\mathsf{PKE}},\sigma) \gets \mathtt{XOF}(\mathit{seed}_{\mathsf{KEM}})$
\STATE \quad $(\mathsf{ek}_{\mathsf{PKE}},\mathsf{dk}_{\mathsf{PKE}}) \gets \textsf{HQC-PKE.KeyGen}(\mathit{seed}_{\mathsf{PKE}})$
\STATE \quad $\mathsf{ek}_{\mathsf{KEM}} \gets \mathsf{ek}_{\mathsf{PKE}}$
\STATE \quad $\mathsf{dk}_{\mathsf{KEM}} \gets (\mathsf{ek}_{\mathsf{KEM}},\mathsf{dk}_{\mathsf{PKE}},\sigma,\mathit{seed}_{\mathsf{KEM}})$
\STATE \quad \textbf{return} $\mathsf{ek}_{\mathsf{KEM}}$, $\mathsf{dk}_{\mathsf{KEM}}$
\vspace{5pt}
\STATE \textsf{HQC-KEM.Encaps}$(\mathsf{ek}_{\mathsf{KEM}})$
\STATE \quad $\mathbf{m} \stackrel{\$}{\gets} \{0,1\}^k$, $\mathit{salt} \stackrel{\$}{\gets} \{0,1\}^{|\mathit{salt}|}$
\STATE \quad $(K,\theta) \gets G(H(\mathsf{ek}_{\mathsf{KEM}}) \,\|\, \mathbf{m} \,\|\, \mathit{salt})$
\STATE \quad $\mathbf{c}_{\mathsf{PKE}} \gets \textsf{HQC-PKE.Encrypt}(\mathsf{ek}_{\mathsf{KEM}},\mathbf{m},\theta)$
\STATE \quad $\mathbf{c}_{\mathsf{KEM}} \gets (\mathbf{c}_{\mathsf{PKE}},\mathit{salt})$
\STATE \quad \textbf{return} $K$, $\mathbf{c}_{\mathsf{KEM}}$
\vspace{5pt}
\STATE \textsf{HQC-KEM.Decaps}$(\mathsf{dk}_{\mathsf{KEM}},\mathbf{c}_{\mathsf{KEM}})$
\STATE \quad Parse $\mathsf{dk}_{\mathsf{KEM}}=(\mathsf{ek}_{\mathsf{KEM}},\mathsf{dk}_{\mathsf{PKE}},\sigma,\mathit{seed}_{\mathsf{KEM}})$
\STATE \quad Parse $\mathbf{c}_{\mathsf{KEM}}=(\mathbf{c}_{\mathsf{PKE}},\mathit{salt})$
\STATE \quad $\mathbf{m}' \gets \textsf{HQC-PKE.Decrypt}(\mathsf{dk}_{\mathsf{PKE}},\mathbf{c}_{\mathsf{PKE}})$
\STATE \quad $(K',\theta') \gets G(H(\mathsf{ek}_{\mathsf{KEM}}) \,\|\, \mathbf{m}' \,\|\, \mathit{salt})$
\STATE \quad $\mathbf{c}'_{\mathsf{PKE}} \gets \textsf{HQC-PKE.Encrypt}(\mathsf{ek}_{\mathsf{KEM}},\mathbf{m}',\theta')$
\STATE \quad $\bar{K} \gets J(H(\mathsf{ek}_{\mathsf{KEM}}) \,\|\, \sigma \,\|\, \mathbf{c}_{\mathsf{KEM}})$
\STATE \quad \textbf{if} $\mathbf{m}'=\bot$ or $\mathbf{c}'_{\mathsf{PKE}}\neq\mathbf{c}_{\mathsf{PKE}}$ \textbf{then} $K' \gets \bar{K}$
\STATE \quad \textbf{return} $K'$
\end{algorithmic}
\end{algorithm}

\begin{figure*}[t]
\centering
\includegraphics[width=0.8\textwidth]{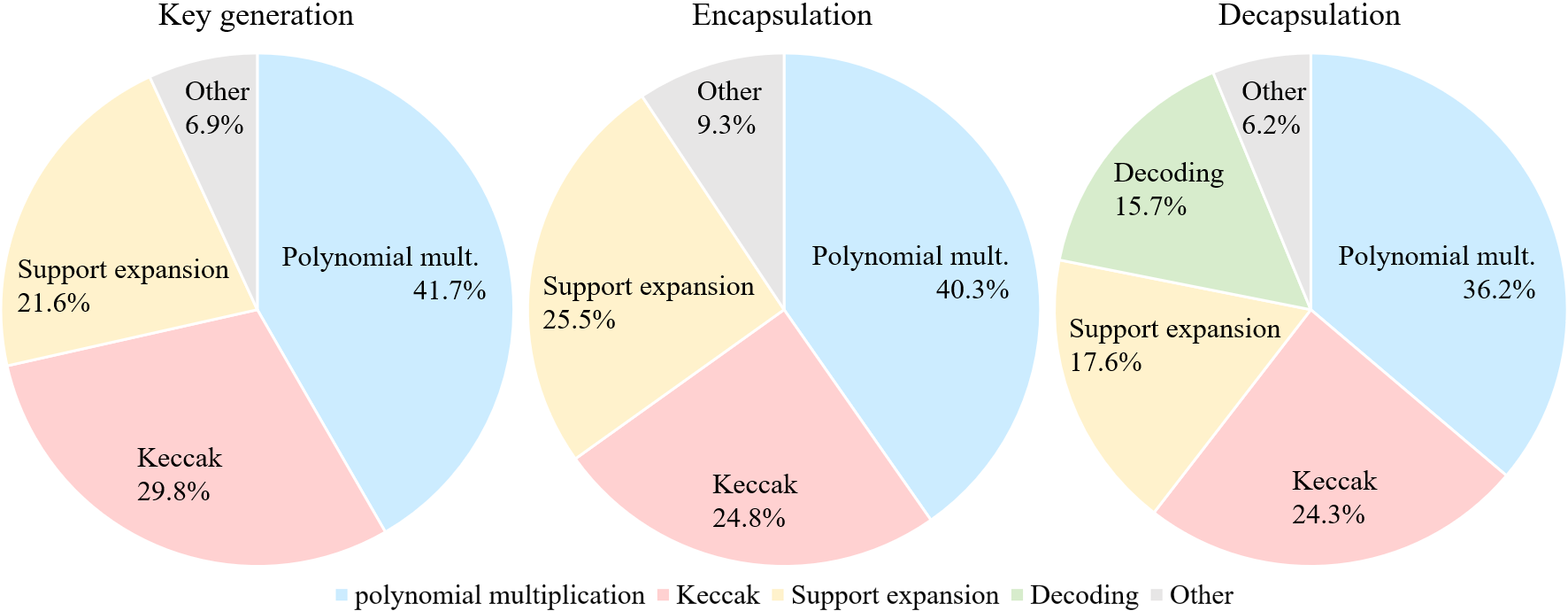}
\caption{Cycle distribution of the Cabral et al. AVX-512 implementation for \HQC{5} on an Intel Core i5-1135G7 processor.}
\label{fig:cabral-profiling}
\end{figure*}

Polynomial multiplication is used to compute $\mathbf{h}\cdot\mathbf{y}$, $\mathbf{h}\cdot\mathbf{r}_2$, $\mathbf{s}\cdot\mathbf{r}_2$, and $\mathbf{u}\cdot\mathbf{y}$. The decoder $\mathcal{C}.\mathsf{Decode}$ invokes Reed-Muller and Reed-Solomon decoding. Figure~\ref{fig:cabral-profiling} shows the contribution of these components to the KEM cycle counts in the AVX-512 implementation of Cabral et al.~\cite{cabral2026accelerating} at \HQC{5}. In the figure, the Keccak category includes the SHA3-256 oracles $H$ and $J$, the SHA3-512 oracles $G$ and $I$, and the SHAKE-256 XOF. Support expansion, the final step of fixed-weight sampling, converts the sampled support positions into a binary vector of length $N$. These results motivate our optimization of polynomial multiplication, Keccak-based functions, decoding, and support expansion.

\subsection{AVX-512 and GFNI}
\label{ssec:avx512-gfni}

AVX-512 extends the x86 SIMD instruction set to 32 512-bit ZMM registers and eight opmask registers. A ZMM register extends the corresponding 256-bit YMM register, which forms its lower half. In this work, we use AVX-512 instructions together with VPCLMULQDQ for binary polynomial multiplication and GFNI for byte-wise arithmetic over $\mathbb{F}_{2^8}^{\mathrm{RS}}$.

The VPCLMULQDQ instruction computes carry-less products of 64-bit operands. The 256-bit form uses YMM registers and computes two 64-bit carry-less products per instruction. The 512-bit form uses ZMM registers and computes four such products.

GFNI provides byte-wise finite-field and affine instructions through \texttt{vgf2p8mulb}, \texttt{vgf2p8affineqb}, and \texttt{vgf2p8affineinvqb}~\cite{IntelIntrinsicsGuide}. The multiplication and inversion instructions are defined over $\mathbb{F}_{2^8}^{\mathrm{AES}}$, whereas the Reed-Solomon code in HQC is defined over $\mathbb{F}_{2^8}^{\mathrm{RS}}$. We use the GFNI affine instructions to map elements from $\mathbb{F}_{2^8}^{\mathrm{RS}}$ to $\mathbb{F}_{2^8}^{\mathrm{AES}}$ before multiplication or inversion and map the results back to $\mathbb{F}_{2^8}^{\mathrm{RS}}$ afterward. Let $T:\mathbb{F}_{2^8}^{\mathrm{RS}}\rightarrow\mathbb{F}_{2^8}^{\mathrm{AES}}$ be the linear isomorphism between the two fields. For $a,b\in\mathbb{F}_{2^8}^{\mathrm{RS}}$, the RS-field multiplication is computed as
\[
a\cdot_{\mathrm{RS}} b = T^{-1}\bigl(T(a)\cdot_{\mathrm{AES}}T(b)\bigr).
\]
The same conversion is used for inversion as $a^{-1}=T^{-1}(T(a)^{-1})$.

Our AVX2 implementation targets processors supporting AVX2, PCLMULQDQ, and GFNI, and our AVX-512 implementation additionally requires AVX-512 and VPCLMULQDQ. The experiments are performed on an Intel Tiger Lake processor. Table~\ref{tab:cpu-coverage} summarizes instruction support on representative x86 microarchitectures.

\begin{table}[t]
\centering
\caption{Instruction support on representative microarchitectures.}
\label{tab:cpu-coverage}
\setlength{\tabcolsep}{3pt}
\begin{tabular}{lcccc}
\toprule
Microarchitecture (year) & AVX2 & AVX-512 & VPCLMULQDQ & GFNI \\
\midrule
Ice Lake (2019) & \checkmark & \checkmark & \checkmark & \checkmark \\
Tiger Lake (2020) & \checkmark & \checkmark & \checkmark & \checkmark \\
Rocket Lake (2021) & \checkmark & \checkmark & \checkmark & \checkmark \\
Alder Lake (2021) & \checkmark & -- & \checkmark & \checkmark \\
Raptor Lake (2022) & \checkmark & -- & \checkmark & \checkmark \\
Meteor Lake (2023) & \checkmark & -- & \checkmark & \checkmark \\
Arrow Lake (2024) & \checkmark & -- & \checkmark & \checkmark \\
Lunar Lake (2024) & \checkmark & -- & \checkmark & \checkmark \\
Skylake-X (2017) & \checkmark & \checkmark & -- & -- \\
Ice Lake Xeon (2021) & \checkmark & \checkmark & \checkmark & \checkmark \\
Sapphire Rapids (2023) & \checkmark & \checkmark & \checkmark & \checkmark \\
Emerald Rapids Xeon (2023) & \checkmark & \checkmark & \checkmark & \checkmark \\
Granite Rapids Xeon (2024) & \checkmark & \checkmark & \checkmark & \checkmark \\
Zen 3 (2020) & \checkmark & -- & -- & -- \\
Zen 4 (2022) & \checkmark & \checkmark & \checkmark & \checkmark \\
Zen 5 (2024) & \checkmark & \checkmark & \checkmark & \checkmark \\
\bottomrule
\end{tabular}
\end{table}

{
\subsection{TLS 1.3 and Post-Quantum TLS}
\label{ssec:pqtls}

TLS 1.3 establishes a shared secret through a key-share exchange in the first round trip. When the key exchange is instantiated with a KEM, the client first generates a KEM key pair and sends the public key in its ClientHello message. The server then encapsulates a shared secret to this public key and returns the resulting ciphertext in its handshake flight. Finally, the client decapsulates the ciphertext with its secret key to recover the same shared secret. The client therefore performs key generation and decapsulation, while the server performs one encapsulation per handshake. The public key and the ciphertext are carried in the handshake flights, so the communication sizes of the KEM add directly to the handshake traffic. Post-quantum KEMs are commonly integrated into TLS through liboqs and oqs-provider~\cite{liboqs}, which expose them as key-share groups to OpenSSL. We use this path in Section~\ref{sec:tls}.
}

\section{Polynomial Multiplication}
\label{sec:multiplication}

This section describes how polynomial multiplication is implemented in our work. The binary polynomial products in the HQC PKE component are computed in $R=\mathbb{F}_2[x]/(x^N-1)$. As shown in Section~\ref{ssec:hqc}, HQC uses four such products, namely $\mathbf{h}\cdot\mathbf{y}$ in key generation, $\mathbf{h}\cdot\mathbf{r}_2$ and $\mathbf{s}\cdot\mathbf{r}_2$ in encryption, and $\mathbf{u}\cdot\mathbf{y}$ in decryption. Since $N$ ranges from $17{,}669$ to $57{,}637$, these products account for a large part of the HQC cycle count.

\subsection{Toom-Cook / Karatsuba for AVX-512}
\label{sssec:mul-tckarat}

Toom-Cook and Karatsuba multiplication decompose a large binary-polynomial product into smaller subproducts and combine them by additions over $\mathbb{F}_2$. In HQC implementations, these decompositions are applied recursively until the subproducts reach a size handled by carry-less multiplication instructions.

For single polynomial multiplication, Robert and V\'eron~\cite{RV23} provide an AVX-512 base multiplication that computes four carry-less products of 64-bit operands per instruction. Their implementation places this base case below fixed Toom-Cook/Karatsuba trees. GFmulOpt of Jang et al.~\cite{Jang26} systematically searches decomposition trees and generates branch-free implementations for AVX2 and Arm Neon. Cabral et al.~\cite{cabral2026accelerating} retain the Toom-Cook/Karatsuba decomposition of the HQC AVX2 implementation. Their lower Karatsuba layer uses both the 256-bit and 512-bit forms of VPCLMULQDQ, but the upper decomposition tree remains at AVX2 width.

We extend GFmulOpt to search and generate Toom-Cook/Karatsuba decomposition trees for AVX-512. We adopt the AVX-512 multiplication of Robert and V\'eron as the base case of the extended framework. With this base multiplication fixed, GFmulOpt selects the sequence of Toom-Cook and Karatsuba decompositions for each HQC parameter set. We use the resulting implementations for $\mathbf{y} \cdot \mathbf{h}$ in key generation and $\mathbf{y} \cdot \mathbf{u}$ in decryption.

In encryption, the two products $\mathbf{r}_2 \cdot \mathbf{h}$ and $\mathbf{r}_2 \cdot \mathbf{s}$ have no data dependency and can be processed in parallel. For this case, we use the \texttt{vect\_mul2x} implementation of Cabral et al.~\cite{cabral2026accelerating} unchanged. It packs $\mathbf{h}$ and $\mathbf{s}$ into the two halves of a ZMM register and duplicates $\mathbf{r}_2$ across both halves, allowing the same Toom-Cook/Karatsuba operations to process both products in parallel.

\subsection{Additive-FFT for AVX-512}
\label{sssec:mul-fafft}

FFT-based multiplication has lower asymptotic complexity than Toom-Cook/Karatsuba multiplication. At the polynomial sizes used in HQC, the transform overhead can offset this advantage. FAFFT reduces the transform cost by using the Frobenius relation. The input polynomials are first converted from the monomial basis to the novel polynomial basis and evaluated over a Frobenius partition $\Sigma$ using addFFT. Since the input polynomials have coefficients in $\mathbb{F}_2$, we have $f(\alpha^2)=f(\alpha)^2$. Hence, the value at $\alpha^{2^i}$ can be derived from the value at $\alpha$ by repeated squaring. FAFFT therefore reduces the number of explicit evaluation points by a factor of $m$ compared with naive additive-FFT multiplication over $\mathbb{F}_{2^m}$.

\begin{figure}[t]
\centering
\includegraphics[width=\columnwidth]{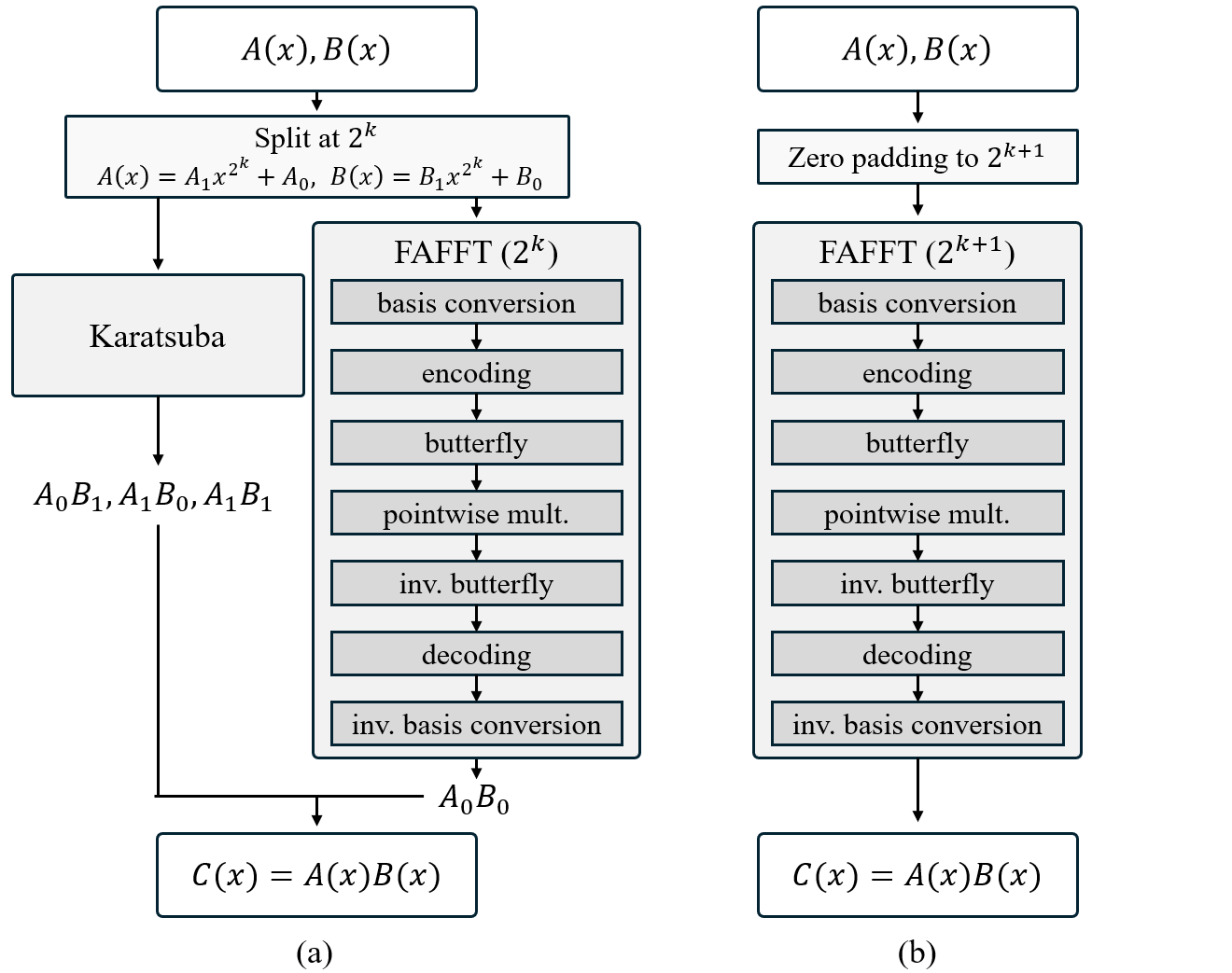}
\caption{FAFFT-based polynomial multiplication of Chen et al.~\cite{cryptoeprint:2026/014} for $N=2^k+r$. (a) Combination of FAFFT and Karatsuba multiplication for \HQC{1} and \HQC{3}. (b) Straightforward FAFFT multiplication for \HQC{5}. The values in parentheses denote the FAFFT input sizes.}
\label{fig:fafft-flow}
\end{figure}

Chen et al.~\cite{cryptoeprint:2026/014} implement FAFFT with AVX2 and GFNI and show that it outperforms Toom-Cook/Karatsuba multiplication at \HQC{5}. Their implementation uses two multiplication structures according to the HQC parameter set. For \HQC{1} and \HQC{3}, the polynomial length has the form $2^k+r$, where $r<2^k$. FAFFT computes the product of the first $2^k$ coefficients, while the products involving the remaining $r$ coefficients are computed separately with Karatsuba multiplication. For \HQC{5}, the input polynomials are padded to $2^{k+1}$ coefficients and the entire product is computed with FAFFT. Each FAFFT computation follows the same sequence of basis conversion, encoding, forward butterfly, pointwise multiplication, inverse butterfly, decoding, and inverse basis conversion. Figure~\ref{fig:fafft-flow} illustrates the two multiplication structures and the FAFFT stages.

We port this FAFFT implementation to AVX-512 and use AVX-512 GFNI for encoding, decoding, and the forward and inverse butterfly operations, thereby increasing the degree of parallelism in the underlying field arithmetic. Specifically, the number of elements processed per iteration increases from 128 to 256 for encoding and decoding, and from 64 to 128 for the butterfly operations.

To further reduce the cost of the butterfly operations, we precompute the twiddle constants used in the fused low stages. Since these constants depend only on the public parameter set and the iteration index, we precompute the constants required for each iteration over 128 elements of $\mathbb{F}_{2^{32}}$ and store them in tables. This replaces the broadcasts and XORs used to construct the constants with table loads. The optimization introduces a trade-off between table storage and instruction count. The tables occupy 11{,}776, 23{,}552, and 47{,}104 bytes for \HQC{1}, \HQC{3}, and \HQC{5}, respectively. We also evaluated this approach for AVX2. However, the AVX2 butterfly processes only one 64-element block per iteration and does not require the two-block packing used in the AVX-512 implementation; replacing constant construction with table loads therefore does not consistently reduce the cycle count. We consequently use the precomputed tables only in the AVX-512 implementation.

We further reduce the transform overhead by eliminating unnecessary basis-conversion operations on the padded zero region. For example, at \HQC{5}, the polynomial occupies only $7{,}205$ bytes of the $8{,}192$-byte FFT core, while the original basis conversion and its inverse process the entire core, including the padded zero region. We derive the corresponding loop ranges from the public parameter set and skip these unnecessary operations in both the AVX2 and AVX-512 implementations. Since the loop ranges depend only on the public parameter set, the resulting control flow and memory-access pattern remain input-independent.

For the Karatsuba component in Figure~\ref{fig:fafft-flow}(a), we use the AVX-512 form of VPCLMULQDQ, which computes four 64-bit carry-less products in parallel.

\subsection{Performance}
\label{ssec:mul-comparison}

Table~\ref{tab:mul-cycles-avx512} reports cycle counts of single polynomial multiplication on a Tiger Lake i5-1135G7. Turbo Boost and Hyper-Threading are disabled, and the benchmark process is pinned to a single core. Each value is the median of the measured samples after warmup, as described in Section~\ref{ssec:eval-setup}. In the table, TC/Karat denotes Toom-Cook/Karatsuba multiplication.

\begin{table}[!h]
\centering
\caption{Cycle counts of polynomial multiplication on AVX-512.}
\label{tab:mul-cycles-avx512}
\begin{tabular}{llrrr}
\toprule
Implementation & Method & \multicolumn{1}{r}{\HQC{1}} & \multicolumn{1}{r}{\HQC{3}} & \multicolumn{1}{r}{\HQC{5}} \\
\midrule
Robert-V\'eron~\cite{RV23}                    & TC/Karat & 15{,}394 & 46{,}604 & 110{,}066 \\
Cabral et al.~\cite{cabral2026accelerating}   & TC/Karat & 17{,}840 & 51{,}217 & 124{,}453 \\
Ours                                    & TC/Karat & 15{,}584 & 46{,}178 & 94{,}539 \\
Ours                                    & FAFFT               & 27{,}650 & 67{,}037 & 90{,}627 \\
\bottomrule
\end{tabular}
\end{table}

The ``TC/Karat'' row of this work is obtained from the GFmulOpt search extended to AVX-512, while the ``FAFFT'' row is the AVX-512 port of the implementation of Chen et al.~\cite{cryptoeprint:2026/014} with the twiddle-table implementation and the reduced basis conversions of Section~\ref{sssec:mul-fafft}. The TC/Karat implementation gives the lowest cycle count at \HQC{1} and \HQC{3}, while the FAFFT implementation gives the lowest cycle count at \HQC{5}. The TC/Karat implementation is close to Robert-V\'eron at \HQC{1} and \HQC{3}, with about 1\% difference, and reduces the cost by 14.1\% at \HQC{5}. Compared with Cabral et al.~\cite{cabral2026accelerating}, the reductions are 12.6\%, 9.8\%, and 24.0\%.

The relative cost of the AVX-512 FAFFT implementation over the Toom-Cook/Karatsuba implementation decreases with the parameter size, from 77\% at \HQC{1} to 45\% at \HQC{3}, and turns into a 4.1\% advantage at \HQC{5}. The crossover is realized by the twiddle-table implementation and the reduced basis conversions of Section~\ref{sssec:mul-fafft}. These changes remove the per-iteration constant construction of the low butterfly stages and the basis-conversion operations on the padded zero part. For AVX-512, TC/Karat gives the lowest cycle count at \HQC{1} and \HQC{3}, while FAFFT gives the lowest cycle count at \HQC{5}. We report both configurations at the KEM level in Section~\ref{ssec:eval-kem}.

The table reports single polynomial multiplication only. For the two encryption products that share the operand $\mathbf{r}_2$, we reuse \texttt{vect\_mul2x} of~\cite{cabral2026accelerating}. Under the same measurement conditions, \texttt{vect\_mul2x} takes 27{,}364, 77{,}041, and 181{,}096 cycles on \HQC{1}, \HQC{3}, and \HQC{5}, respectively. These values are below twice the corresponding single polynomial multiplication cost, showing the benefit of shared-operand packing.

\section{RM/RS Decoding}
\label{sec:decoding}

This section presents our optimization of the Reed-Muller and Reed-Solomon decoding procedures in HQC. HQC uses a concatenated code that combines an outer Reed-Solomon code over $\mathbb{F}_{2^8}^{\mathrm{RS}}$ and an inner Reed-Muller code $\mathrm{RM}(1,7)$ of length 128. The decoder operates in two stages. The Reed-Muller stage performs a maximum-likelihood decoding of each 128-bit block via the Fast Hadamard Transform, producing one byte of the outer codeword per block. The Reed-Solomon stage then runs an algebraic decoder over $\mathbb{F}_{2^8}^{\mathrm{RS}}$ to correct residual errors and recover the message. The two stages have very different arithmetic profiles. The Reed-Muller stage mainly operates on 16-bit integer values, using additions and subtractions in the Hadamard transform and signed comparisons in \texttt{find\_peaks}. The Reed-Solomon stage mainly uses $\mathbb{F}_{2^8}^{\mathrm{RS}}$ multiplications and inversions. We therefore present them separately in Sections~\ref{ssec:rm} and~\ref{ssec:rs}.

\subsection{Reed-Muller Decoder}
\label{ssec:rm}

\subsubsection{Decoder phases}
\label{sssec:rm-phases}

The Reed-Muller decoder recovers one byte from \texttt{MULTIPLICITY} duplicated 128-bit codeword copies. It consists of three phases, which we refer to as \texttt{expand\_and\_sum}, \texttt{hadamard}, and \texttt{find\_peaks}, following the naming convention of the HQC AVX2 implementation. The \texttt{expand\_and\_sum} phase accumulates the duplicated codeword copies into 128 signed 16-bit values. The \texttt{hadamard} phase applies a seven-stage Fast Hadamard Transform in place. After correcting the zero coefficient by $64 \cdot \texttt{MULTIPLICITY}$, the \texttt{find\_peaks} phase locates the entry of maximum absolute value in the transformed vector. The decoded byte is obtained from the corresponding 7-bit position and sign bit.

\subsubsection{Prior implementations}
\label{sssec:rm-prior}

The HQC NIST submission~\cite{NISTPQC-R4:HQC22} implements the three Reed-Muller phases with AVX2. The \texttt{hadamard} phase uses \texttt{\_mm256\_hadd\_epi16} and \texttt{\_mm256\_hsub\_epi16}, which compute horizontal additions and subtractions on 16-bit lanes. To find the maximum, the \texttt{find\_peaks} phase performs a binary search with a fixed number of iterations over the possible coefficient values. It then obtains the position of the maximum with \texttt{\_bit\_scan\_forward}, which returns the index of the least significant set bit.

Chen et al.~\cite{cryptoeprint:2026/014} optimize \texttt{expand\_and\_sum} with AVX2. They expand each 16-bit input word into 16 16-bit values using vector comparisons and accumulate the duplicated codeword copies with \texttt{vpaddw}. Their \texttt{hadamard} and \texttt{find\_peaks} follow the HQC AVX2 implementation.

Cabral et al.~\cite{cabral2026accelerating} reimplement the three phases with AVX-512. For \texttt{expand\_and\_sum}, they replace the AVX2 bit-extraction pattern with \texttt{vpmovm2w} and accumulate the duplicated codeword copies with \texttt{vpaddw}. For \texttt{hadamard}, they unroll the seven butterfly stages and use \texttt{\_mm512\_mask\_blend\_epi16} with byte shifts and permutations. For \texttt{find\_peaks}, they keep the binary-search structure of the reference and replace the AVX2 extraction step with a constant-time mask-based extraction of the maximum index.

\subsubsection{Our Reed-Muller decoder}
\label{sssec:rm-ours}

We modify the \texttt{find\_peaks} phase for both AVX2 and AVX-512. We use the \texttt{expand\_and\_sum} and \texttt{hadamard} implementations of Cabral et al.~\cite{cabral2026accelerating} for AVX-512 and those of Chen et al.~\cite{cryptoeprint:2026/014} for AVX2. In both settings, we replace the previous \texttt{find\_peaks} with the maximum-search method described below.

\begin{figure}[t]
\centering
\includegraphics[width=\columnwidth]{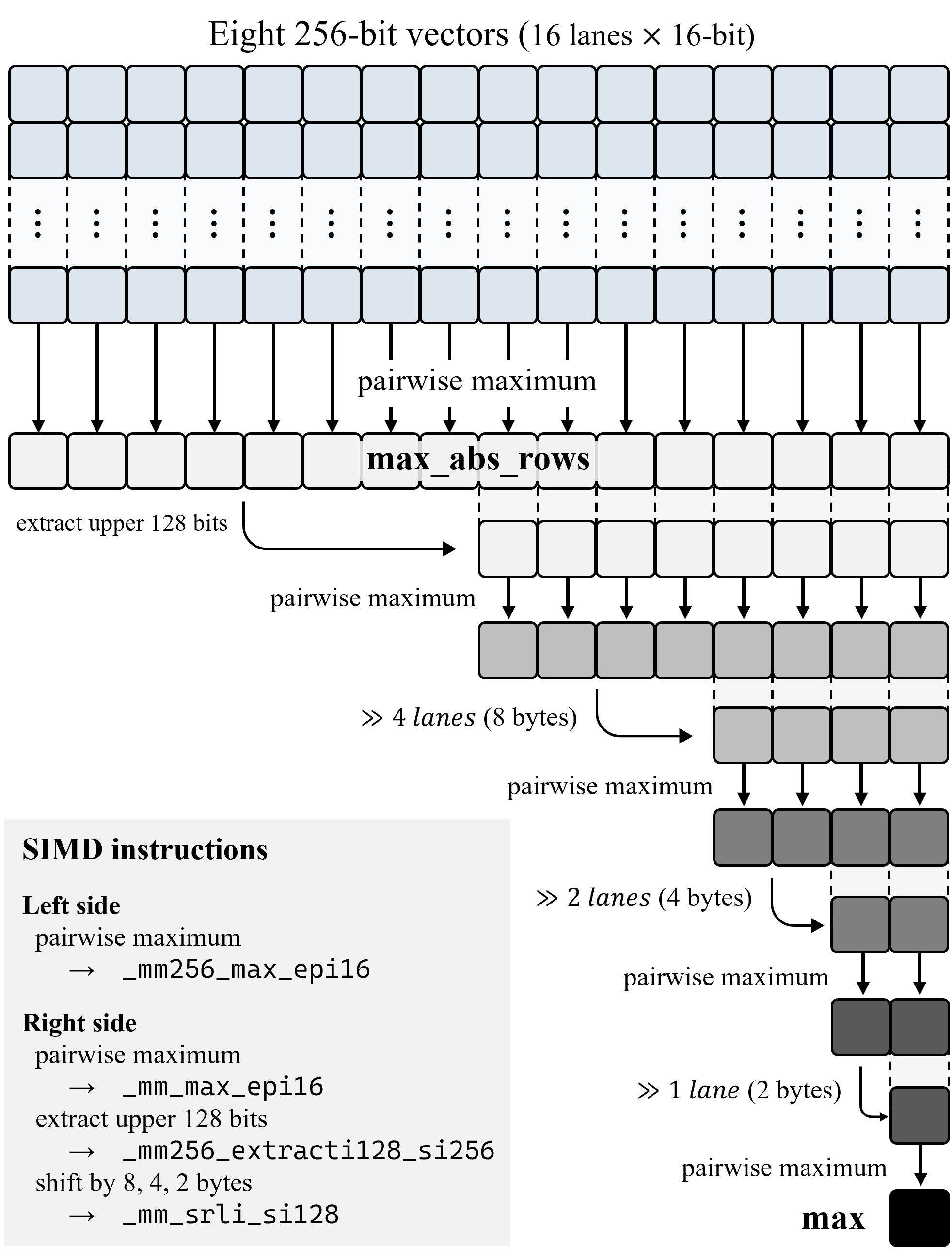}
\caption{Maximum reduction in the AVX2 implementation of \texttt{find\_peaks}.}
\label{fig:find-peaks-reduction}
\end{figure}

Both \texttt{find\_peaks} implementations retain the compact 256-bit vector layout and replace the binary search for the maximum with a horizontal vector reduction. Figure~\ref{fig:find-peaks-reduction} shows this reduction in our AVX2 implementation. It first computes lane-wise maxima over eight vectors and then reduces the 16 remaining values to one maximum. The AVX-512 implementation performs the same sequence on 256-bit vectors. Once the maximum is obtained, comparison masks identify its row. The mask of the selected row is collapsed, and \texttt{\_bit\_scan\_forward} returns the column index. The AVX2 implementation selects the row and its coefficient with constant-time masks. The AVX-512 implementation uses comparison masks for the row and \texttt{\_mm256\_permutexvar\_epi16} for the coefficient. This avoids a data-dependent scalar index into a temporary array. The sign bit of the final output is then derived from the selected coefficient without data-dependent branches.

The resulting \texttt{find\_peaks} has lower cycle counts than both the HQC AVX2 implementation and the AVX-512 implementation of Cabral et al., as shown in Section~\ref{ssec:decoder-comparison}.

\subsection{Reed-Solomon Decoder}
\label{ssec:rs}

\subsubsection{Decoder phases}
\label{sssec:rs-phases}

The Reed-Solomon decoder corrects up to $\delta$ byte errors in the outer codeword by computing syndromes, recovering the error-locator polynomial, locating the error positions, and applying Forney's formula. It consists of six phases. The \texttt{compute\_syndromes} phase evaluates the received word at the $2\delta$ syndrome points in $\mathbb{F}_{2^8}^{\mathrm{RS}}$. The \texttt{compute\_elp} phase runs the Berlekamp-Massey algorithm to recover the error-locator polynomial $\sigma$. The \texttt{compute\_roots} phase locates the error positions from the roots of $\sigma$. The \texttt{compute\_z\_poly} phase forms the auxiliary polynomial $z$ used by Forney's formula. The \texttt{compute\_error\_values} phase applies Forney's formula to compute the nonzero coefficients of the error polynomial. The \texttt{correct\_errors} phase XORs the error pattern into the received word. The first five phases mainly consist of $\mathbb{F}_{2^8}^{\mathrm{RS}}$ arithmetic, while \texttt{correct\_errors} is a byte-wise XOR step.

\subsubsection{Prior implementations}
\label{sssec:rs-prior}

The optimized AVX2 implementation in the HQC NIST submission~\cite{NISTPQC-R4:HQC22} evaluates syndromes with the AVX2 function \texttt{gf\_mul\_vect}, which implements packed $\mathbb{F}_{2^8}^{\mathrm{RS}}$ multiplication using PCLMULQDQ. Several other arithmetic paths invoke the scalar \texttt{gf\_mul} function, which computes one carry-less product with PCLMULQDQ followed by polynomial reduction.

Chen et al.~\cite{cryptoeprint:2026/014} also optimize $\mathbb{F}_{2^8}^{\mathrm{RS}}$ multiplication on AVX2+GFNI. In the Reed-Solomon decoder, their syndrome computation converts bytes to the AES field representation with \texttt{vgf2p8affineqb}, uses \texttt{vgf2p8mulb} for packed products, and converts the result back to the Reed-Solomon representation. The remaining decoding phases still rely on scalar \texttt{gf\_mul} calls.

Cabral et al.~\cite{cabral2026accelerating} mainly introduce AVX-512 in \texttt{compute\_error\_values} of the Reed-Solomon decoder. This phase uses Montgomery batch inversion and 512-bit packed field multiplication with the AVX-512 form of VPCLMULQDQ without GFNI. The remaining decoding phases keep the HQC AVX2 field-arithmetic structure.

\subsubsection{Our Reed-Solomon decoder}
\label{sssec:rs-ours}

GFNI provides \texttt{vgf2p8mulb} for packed byte multiplication and \texttt{vgf2p8affineinvqb} for packed byte inversion over $\mathbb{F}_{2^8}^{\mathrm{AES}}$. Since the Reed-Solomon code in HQC is defined over $\mathbb{F}_{2^8}^{\mathrm{RS}}$, we use \texttt{vgf2p8affineqb} to convert the byte representation before and after each GFNI field operation. Using this conversion, we replace the PCLMULQDQ-based field arithmetic in the five $\mathbb{F}_{2^8}^{\mathrm{RS}}$-arithmetic phases with packed GFNI operations for both AVX2 and AVX-512. Since each vector lane stores one byte, a 256-bit GFNI instruction processes 32 field elements and a 512-bit GFNI instruction processes 64 field elements. Our AVX2 decoder uses 256-bit vectors in all five phases and evaluates 32 Chien search candidates per iteration. Our AVX-512 decoder uses 256-bit vectors for the syndrome, error-locator, auxiliary polynomial, and error-value computations. Its Chien search uses 512-bit vectors to evaluate 64 candidates per iteration. The phases are optimized as follows.

\begin{figure}[h!]
\centering
\includegraphics[width=\columnwidth]{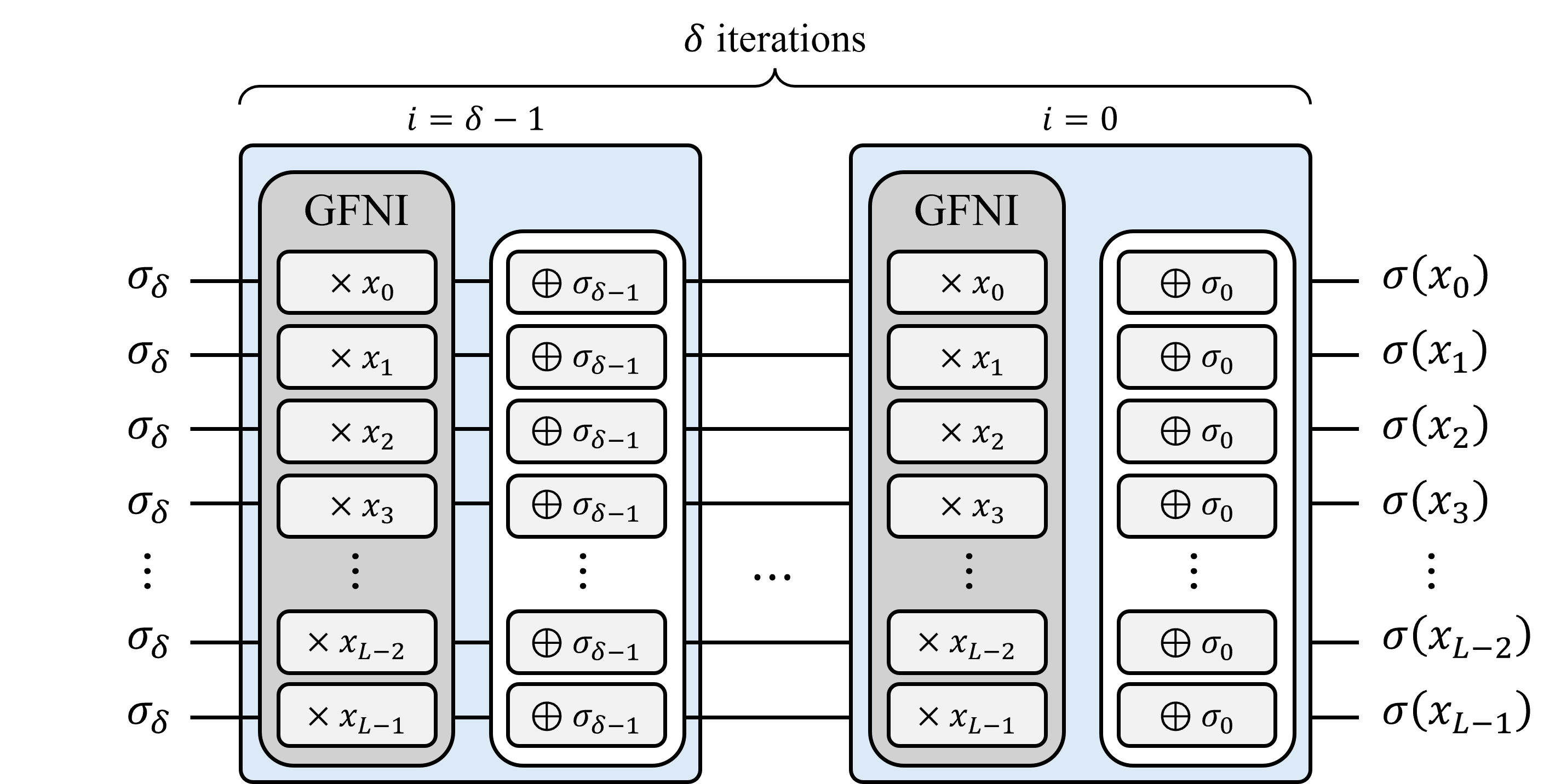}
\caption{Parallel Chien search based on Horner evaluation, where $L=32$ for AVX2 and $L=64$ for AVX-512.}
\label{fig:chien-search}
\end{figure}

\begin{itemize}
\item \emph{\texttt{compute\_syndromes}.} This phase evaluates the received codeword at all $2\delta$ syndrome points. We process the evaluations in parallel with packed Horner computation, where the $\mathbb{F}_{2^8}^{\mathrm{RS}}$ multiplications use \texttt{vgf2p8mulb} and the additions are XORs.

\item \emph{\texttt{compute\_elp}.} This phase runs the Berlekamp-Massey update for the error-locator polynomial. We retain the control flow of the HQC AVX2 implementation and replace the inner $\mathbb{F}_{2^8}^{\mathrm{RS}}$ multiplications with \texttt{vgf2p8mulb}. The discrepancy inversion processes one field element at a time using \texttt{vgf2p8affineinvqb}.

\item \emph{\texttt{compute\_roots}.} This phase locates the roots of the error-locator polynomial. Prior implementations~\cite{NISTPQC-R4:HQC22,cryptoeprint:2026/014,cabral2026accelerating} use an additive-FFT-based procedure for root finding. This approach reduces the number of field evaluations, but it also requires basis conversions and data movement. These operations are not well matched to GFNI, which is most effective when the same byte-field operation is applied to many lanes. Since the candidates lie in $\mathbb{F}_{2^8}^{\mathrm{RS}}$ and the degree of $\sigma$ is bounded by $\delta$, we use Chien search and evaluate $\sigma(x)$ over the nonzero field elements in parallel. Each iteration evaluates 64 candidates by Horner evaluation with \texttt{vgf2p8mulb} and XOR, and the roots are found by comparing the results with zero. Figure~\ref{fig:chien-search} illustrates this packed evaluation.

\item \emph{\texttt{compute\_z\_poly}.} This phase forms the auxiliary polynomial $z$ from $\sigma$ and the truncated syndrome sequence. We compute the convolution in vector blocks, where each accumulation step is a \texttt{vgf2p8mulb} followed by an XOR.

\item \emph{\texttt{compute\_error\_values}.} This phase evaluates Forney's formula at the error locations. We collect the error locations into byte vectors, compute the required inverses with \texttt{vgf2p8affineinvqb}, and evaluate the numerator and denominator in parallel with \texttt{vgf2p8mulb}.
\end{itemize}

The resulting Reed-Solomon decoder replaces the PCLMULQDQ-based field arithmetic in the critical loops with packed GFNI operations. Cycle counts for the individual phases and for the full \texttt{reed\_solomon\_decode} are shown in Section~\ref{ssec:decoder-comparison}.

\subsection{Performance}
\label{ssec:decoder-comparison}

Table~\ref{tab:decoder-cycles} reports cycle counts for the Reed-Muller and Reed-Solomon decoders on a Tiger Lake i5-1135G7. Turbo Boost and Hyper-Threading are disabled, and the benchmark process is pinned to a single core. The measurements follow the same protocol as Table~\ref{tab:mul-cycles-avx512}. For both decoders, phase rows are measured with internal timing code placed around the named phases, and the ``Total clean'' row is measured around the complete decoder call in a clean build without this timing code. Hence, the phase rows indicate where the reductions occur, while ``Total clean'' is the decoder cost used for comparison. We group the results by AVX2 and AVX-512 implementations, and mark implementations using GFNI with $*$. The AVX2 setting compares the HQC AVX2 implementation~\cite{NISTPQC-R4:HQC22}, Chen et al.~\cite{cryptoeprint:2026/014}, and our AVX2+GFNI build. The AVX-512 setting compares the AVX-512 implementation of Cabral et al.~\cite{cabral2026accelerating} against our AVX-512+GFNI build.

For the Reed-Muller decoder, our implementation of Section~\ref{sssec:rm-ours} reduces the clean per-call cost of \texttt{reed\_muller\_decode} by 67.9\%--77.9\% relative to the HQC AVX2 implementation, by 18.2\%--20.4\% relative to Chen et al. on AVX2, and by 3.7\%--9.9\% relative to Cabral et al. on AVX-512. Compared with the prior optimized implementations, the reduction mainly comes from \texttt{find\_peaks}. The \texttt{find\_peaks} phase is reduced by 25.7\%--29.7\% over Chen et al. on AVX2 and by 20.5\%--25.7\% over Cabral et al. on AVX-512. The AVX2 \texttt{hadamard} phase follows the same Fast Hadamard Transform structure as the HQC AVX2 implementation, based on \texttt{hadd} and \texttt{hsub}. Its profiled phase row is higher than that of the HQC AVX2 implementation because the phase counter places fence and timestamp boundaries around the transform. Therefore, this row is used only to locate where the costs are spent, while the clean full-decoder row is used for decoder-cost comparison. As a result, the improvement in \texttt{find\_peaks} is reflected in the full Reed-Muller decoding cost.

For the Reed-Solomon decoder, the GFNI-based reimplementation of Section~\ref{sssec:rs-ours} reduces the clean per-call cost of \texttt{reed\_solomon\_decode} by 84.7\%--88.8\% relative to the HQC AVX2 implementation, by 83.2\%--87.1\% relative to Chen et al. on AVX2, and by 88.9\%--91.0\% relative to Cabral et al. on AVX-512. The phase-level cycle counts of Chen et al. show a large reduction in \texttt{compute\_syndromes}, where their implementation uses GFNI, but the other $\mathbb{F}_{2^8}^{\mathrm{RS}}$ arithmetic phases remain close to the HQC AVX2 implementation. These results show that our gain is concentrated in the five phases that perform $\mathbb{F}_{2^8}^{\mathrm{RS}}$ arithmetic. The final byte-wise XOR phase \texttt{correct\_errors} runs under 100 cycles in every configuration and is omitted from the table.

\begin{table*}[!h]
\centering
\begin{threeparttable}
\caption{Cycle counts of Reed-Muller and Reed-Solomon decoding. Phase rows are measured with internal timing code. Total clean rows are measured around complete decoder calls without this timing code.}
\label{tab:decoder-cycles}
\begin{tabular}{l l r r r l r r r}
\toprule
& \multicolumn{4}{c}{AVX2} & \multicolumn{4}{c}{AVX-512} \\
\cmidrule(lr){2-5} \cmidrule(lr){6-9}
Phase & Impl. & \multicolumn{1}{r}{\HQC{1}} & \multicolumn{1}{r}{\HQC{3}} & \multicolumn{1}{r}{\HQC{5}} & Impl. & \multicolumn{1}{r}{\HQC{1}} & \multicolumn{1}{r}{\HQC{3}} & \multicolumn{1}{r}{\HQC{5}} \\
\midrule
\multicolumn{9}{c}{\textit{Reed-Muller decoder}} \\
\midrule
\texttt{expand\_and\_sum} & \cite{NISTPQC-R4:HQC22} & 30{,}178 & 66{,}556 & 108{,}318 & \cite{cabral2026accelerating} & 2{,}558 & 4{,}195 & 6{,}551 \\
                          & \cite{cryptoeprint:2026/014} & 2{,}004 & 2{,}699 & 4{,}293 & Ours & 2{,}561 & 3{,}964 & 6{,}458 \\
                          & Ours                 & 2{,}003 & 2{,}671 & 4{,}368 & & & & \\
\midrule
\texttt{hadamard}         & \cite{NISTPQC-R4:HQC22} & 5{,}773 & 6{,}942 & 11{,}098 & \cite{cabral2026accelerating} & 4{,}651 & 5{,}671 & 8{,}748 \\
                          & \cite{cryptoeprint:2026/014} & 7{,}711 & 10{,}611 & 17{,}093 & Ours & 4{,}469 & 5{,}538 & 9{,}374 \\
                          & Ours                 & 7{,}486 & 10{,}847 & 16{,}911 & & & & \\
\midrule
\texttt{find\_peaks}      & \cite{NISTPQC-R4:HQC22} & 13{,}114 & 16{,}763 & 26{,}950 & \cite{cabral2026accelerating} & 9{,}934 & 12{,}802 & 20{,}821 \\
                          & \cite{cryptoeprint:2026/014} & 12{,}947 & 16{,}535 & 27{,}040 & Ours & 7{,}902 & 9{,}734 & 15{,}461 \\
                          & Ours                 & 9{,}623 & 11{,}824 & 19{,}005 & & & & \\
\midrule
Total clean & \cite{NISTPQC-R4:HQC22} & 45{,}067 & 86{,}073 & 137{,}703 & \cite{cabral2026accelerating} & 11{,}128 & 15{,}326 & 24{,}569 \\
      & \cite{cryptoeprint:2026/014} & 17{,}696 & 23{,}899 & 38{,}237 & Ours & 10{,}714 & 13{,}813 & 22{,}403 \\
      & Ours                 & 14{,}475 & 19{,}087 & 30{,}431 & & & & \\
\midrule
\midrule
\multicolumn{9}{c}{\textit{Reed-Solomon decoder}} \\
\midrule
\texttt{syndromes}    & \cite{NISTPQC-R4:HQC22} & 5{,}597 & 6{,}822 & 21{,}985 & \cite{cabral2026accelerating} & 3{,}579 & 4{,}349 & 13{,}881 \\
                      & \cite{cryptoeprint:2026/014}$^*$ & 211 & 240 & 490 & Ours$^*$ & 128 & 138 & 315 \\
                      & Ours$^*$             & 183 & 208 & 315 & & & & \\
\midrule
\texttt{elp}          & \cite{NISTPQC-R4:HQC22} & 17{,}937 & 19{,}709 & 47{,}069 & \cite{cabral2026accelerating} & 17{,}738 & 19{,}622 & 46{,}712 \\
                      & \cite{cryptoeprint:2026/014} & 17{,}783 & 19{,}790 & 47{,}061 & Ours$^*$ & 3{,}089 & 3{,}298 & 5{,}954 \\
                      & Ours$^*$             & 7{,}505 & 8{,}024 & 14{,}535 & & & & \\
\midrule
\texttt{roots}        & \cite{NISTPQC-R4:HQC22} & 9{,}401 & 17{,}666 & 18{,}195 & \cite{cabral2026accelerating} & 9{,}662 & 17{,}025 & 18{,}016 \\
                      & \cite{cryptoeprint:2026/014} & 10{,}784 & 16{,}181 & 18{,}179 & Ours$^*$ & 392 & 420 & 877 \\
                      & Ours$^*$             & 677 & 730 & 1{,}604 & & & & \\
\midrule
\texttt{z\_poly}      & \cite{NISTPQC-R4:HQC22} & 1{,}174 & 1{,}323 & 4{,}219 & \cite{cabral2026accelerating} & 1{,}189 & 1{,}323 & 4{,}246 \\
                      & \cite{cryptoeprint:2026/014} & 1{,}195 & 1{,}323 & 4{,}208 & Ours$^*$ & 391 & 411 & 695 \\
                      & Ours$^*$             & 417 & 403 & 697 & & & & \\
\midrule
\texttt{err\_values}  & \cite{NISTPQC-R4:HQC22} & 26{,}596 & 29{,}024 & 76{,}508 & \cite{cabral2026accelerating} & 10{,}399 & 12{,}083 & 23{,}299 \\
                      & \cite{cryptoeprint:2026/014} & 26{,}635 & 28{,}949 & 76{,}484 & Ours$^*$ & 959 & 1{,}114 & 1{,}804 \\
                      & Ours$^*$             & 984 & 1{,}114 & 1{,}851 & & & & \\
\midrule
Total clean & \cite{NISTPQC-R4:HQC22} & 62{,}533 & 75{,}294 & 169{,}708 & \cite{cabral2026accelerating} & 44{,}101 & 55{,}398 & 107{,}107 \\
      & \cite{cryptoeprint:2026/014}$^*$ & 56{,}922 & 67{,}463 & 147{,}810 & Ours$^*$ & 4{,}904 & 5{,}246 & 9{,}654 \\
      & Ours$^*$             & 9{,}580 & 10{,}321 & 19{,}075 & & & & \\
\bottomrule
\end{tabular}%
\begin{tablenotes}[flushleft]
\footnotesize
\item[$*$] Uses the GFNI instruction set.
\end{tablenotes}
\end{threeparttable}
\end{table*}

\section{Keccak-Based Functions and Fixed-Weight Sampling}
\label{sec:keccak}

All symmetric primitives used by HQC are built on the Keccak-$f[1600]$ permutation. In this section, we describe the AVX-512 implementation of the SHA3-512 oracles and the AVX-512 support expansion used in fixed-weight sampling.

\subsection{Background}
\label{ssec:keccak-background}

HQC instantiates four random oracles together with a SHAKE-based extendable-output function (XOF) used for vector expansion and rejection sampling. All five primitives use the Keccak-$f[1600]$ permutation and differ in the absorb pattern, the output length, and the domain separator. We list the call sites below.

\begin{itemize}
\item \emph{Hash oracles $H$ and $J$} are instantiated with SHA3-256. The oracle $H$ takes the encapsulation key $ek$ (between $2{,}241$ and $7{,}245$ bytes depending on the parameter set) as input. The oracle $J$ takes $H(ek)$, $\sigma$, the ciphertext, and the salt as input. Both oracles produce a 32-byte output. The oracle $H$ is called once in encapsulation and once in decapsulation, while $J$ is called once in decapsulation to derive the implicit-rejection key.
\item \emph{Hash oracles $G$ and $I$} are instantiated with SHA3-512. The oracle $G$ derives the per-encapsulation seed $(K, \theta)$ from the message and salt, and is called once per encapsulation and once per decapsulation. The oracle $I$ derives the key-pair seed from a single 32-byte input and is called once per key generation.
\item \emph{SHAKE-256 expansion} produces the public vector $\mathbf{h}$ from a 32-byte seed during both key generation and encapsulation, and additionally drives the rejection-sampling XOF used for the low-weight vector samplers in all three KEM operations.
\item \emph{Fixed-weight sampling} generates low-weight vectors in $R_\omega$, $R_{\omega_r}$, and $R_{\omega_e}$ from SHAKE-256 output. The sampler first obtains support positions by rejection sampling and then expands the support into a binary vector in $R$. This final support-expansion step is measured separately in Section~\ref{ssec:keccak-comparison}.
\end{itemize}

\subsection{Prior Implementations}
\label{ssec:keccak-prior}

The HQC AVX2 implementation distributed with the NIST submission~\cite{NISTPQC-R4:HQC22} uses the scalar Keccak implementation from PQClean~\cite{SSR:KSSW22} without SIMD instructions.

Cabral et al.~\cite{cabral2026accelerating} introduced a dedicated AVX-512 implementation of Keccak-$f[1600]$ that maps the 1600-bit state across five 512-bit registers, with five active 64-bit lanes per register. The 24 permutation rounds are unrolled and rely on \texttt{\_mm512\_ternarylogic\_epi64} for the three-way XORs of the $\theta$ step, on \texttt{\_mm512\_rolv\_epi64} for the per-lane rotations of the $\rho$ step, and on \texttt{\_mm512\_permutexvar\_epi64} for the cross-lane shuffles of the $\pi$ step. On top of this permutation, Cabral et al. inline absorb-and-squeeze wrappers for the two SHA3-256 oracles $H$ and $J$ and provide \texttt{shake256\_512}, which is specialized to the expansion of the public vector $\mathbf{h}$. These implementations are used only on the encapsulation and decapsulation paths. The two SHA3-512 oracles $G$ and $I$ and the incremental SHAKE-256 used by rejection sampling remain on the scalar PQClean wrappers in their implementation.

The \HQC{3} implementation accompanying Cabral et al.~\cite{cabral2026accelerating} absorbed only 33 of the 67 input blocks of the oracle $J$. Since $J$ is used only to derive the implicit-rejection key, this error is not detected by known-answer tests based on valid ciphertexts. A subsequent update corrected the absorption. The correction adds 34 Keccak-$f[1600]$ permutations to \HQC{3} decapsulation. We use the corrected implementation for both the Cabral et al. baseline and our implementations.

The XKCP project~\cite{XKCP} provides a third-party AVX-512 implementation of Keccak-$f[1600]$ with the same byte-level state layout as the PQClean scalar code, a 25-element array of \texttt{uint64\_t}. We use this implementation in place of the scalar permutation called by the PQClean wrappers.

\subsection{Our Keccak Strategy}
\label{ssec:keccak-ours}

\subsubsection{SHA3 oracles}
\label{sssec:keccak-hashes}

We use the AVX-512 implementations of $H$ and $J$ from Cabral et al.~\cite{cabral2026accelerating}. For the two SHA3-512 oracles $G$ and $I$, which Cabral et al. leave on scalar wrappers, we provide AVX-512 implementations that follow the same template as their $H$ and $J$. The state uses the same five-register layout. The absorb phase loads the input into the first nine lanes of the state with two masked loads, applies the SHA3-512 padding and domain separator at compile-determined byte offsets, and calls the Cabral et al. permutation implementation once or twice depending on the input length. The squeeze phase stores the 64-byte output from the rate region of the state with two masked stores. The number of permutation calls is fixed for each parameter set. The oracle $I$ uses one permutation for all parameter sets, while $G$ uses one at \HQC{1} and two at \HQC{3} and \HQC{5}.

\subsubsection{Fixed-weight sampling}
\label{sssec:keccak-sampling}

Fixed-weight sampling in HQC consists of support generation from SHAKE-256 followed by support expansion into a binary vector. For SHAKE-256, we replace the scalar Keccak-$f[1600]$ permutation with the XKCP AVX-512 implementation~\cite{XKCP}. Rejection sampling ends with \texttt{vect\_write\_support\_to\_vector}, which expands the sampled $\omega$ support positions into a binary vector of length $N$. In a constant-time implementation, the sampled positions are not used as memory addresses. Therefore, support expansion scans all output words against all support positions and accumulates the matching bits with branch-free masks. Since the inner loop is repeated for each output word and support position, this step increases the cost of fixed-weight sampling. The HQC AVX2 implementation and Cabral et al.~\cite{cabral2026accelerating} both use the AVX2 implementation of this function, which compares four 64-bit output words at a time with a \texttt{vpcmpeqq}-and-AND pattern. We reduce this cost by processing eight 64-bit output words per iteration with AVX-512, using \texttt{\_mm512\_cmpeq\_epi64\_mask} and mask-controlled lane selection.

\subsection{Performance}
\label{ssec:keccak-comparison}

Table~\ref{tab:keccak-cycles} reports the cycle counts of the SHA3-512 oracles and support expansion optimized in this work. The two SHA3-512 oracles $G$ and $I$ are compared against Cabral et al.~\cite{cabral2026accelerating}, whose implementation keeps these oracles on scalar wrappers. The support-expansion function \texttt{vect\_write\_support\_to\_vector} is also compared against Cabral et al., who use the AVX2 implementation for this function. Our AVX-512 implementations reduce the per-call cost by 37.4\%--46.4\% for $G$, by 46.2\%--46.7\% for $I$, and by 34.5\%--36.8\% for support expansion.

\begin{table}[t]
\centering
\caption{Cycle counts of SHA3-512 oracles and support expansion. Support expansion denotes \texttt{vect\_write\_support\_to\_vector}.}
\label{tab:keccak-cycles}
\begin{tabular}{llrrr}
\toprule
Function & Implementation & \multicolumn{1}{r}{\HQC{1}} & \multicolumn{1}{r}{\HQC{3}} & \multicolumn{1}{r}{\HQC{5}} \\
\midrule
$G$ (\texttt{hash\_g}) & \cite{cabral2026accelerating} & 1{,}646 & 3{,}206 & 2{,}769 \\
                       & Ours                        & 913   & 1{,}719 & 1{,}734 \\
\midrule
$I$ (\texttt{hash\_i}) & \cite{cabral2026accelerating} & 1{,}729 & 1{,}712 & 1{,}699 \\
                       & Ours                        & 921   & 913   & 914   \\
\midrule
Support expansion
                       & \cite{cabral2026accelerating} & 5{,}778 & 17{,}175 & 35{,}607 \\
                       & Ours                                 & 3{,}782 & 10{,}951 & 22{,}490 \\
\bottomrule
\end{tabular}
\end{table}

The two SHA3-256 oracles $H$ and $J$ and the \texttt{shake256\_512} implementation are taken from Cabral et al.~\cite{cabral2026accelerating}. We also call \texttt{shake256\_512} in key generation to generate the public vector $\mathbf{h}$. Since this change is reflected in the key-generation cycle count, we discuss it in the end-to-end evaluation.

\section{Experimental Evaluation}
\label{sec:evaluation}

This section reports the cycle counts of our HQC implementations and compares them with the AVX2 implementation distributed with the HQC NIST submission~\cite{NISTPQC-R4:HQC22} and the implementations of Jang et al.~\cite{Jang26}, Chen et al.~\cite{cryptoeprint:2026/014}, Robert and V\'eron~\cite{RV23}, and Cabral et al.~\cite{cabral2026accelerating}. We verify our implementation against the HQC KAT files, which contain 100 test vectors for each parameter set. For each of \HQC{1}, \HQC{3}, and \HQC{5}, our \texttt{crypto\_kem\_keypair}, \texttt{crypto\_kem\_enc}, and \texttt{crypto\_kem\_dec} functions reproduce the reference public key, secret key, ciphertext, and shared secret for all test vectors. Our implementation does not introduce secret-dependent branches or memory accesses. Its loop ranges and execution paths depend only on public parameters.

\subsection{Setup}
\label{ssec:eval-setup}

All measurements are taken on a single core of an Intel Core i5-1135G7 (Tiger Lake) clocked at $2.4$~GHz, with Turbo Boost disabled through the operating system power management, Hyper-Threading disabled, and the benchmark process pinned to a single physical core. This setting avoids frequency changes during benchmarking and makes the cycle counts comparable across runs. Cycle counts are obtained through the \texttt{rdtscp} instruction with leading and trailing \texttt{lfence} barriers placed around each measurement window. Each reported value is the median of 1{,}000 measured samples after 10{,}000 warmup iterations. Each sample averages 100 iterations, and the warmup samples are excluded before aggregation. All binaries are built with \texttt{gcc} 12 using \texttt{-O3}, \texttt{-funroll-all-loops}, and \texttt{-flto}. The main flags for the AVX2 builds are \texttt{-mavx2} and \texttt{-mpclmul}, while the AVX2+GFNI builds additionally use \texttt{-mgfni}. The AVX-512 builds use these AVX2 flags together with \texttt{-mavx512f}, \texttt{-mavx512bw}, \texttt{-mavx512dq}, \texttt{-mavx512vl}, and \texttt{-mvpclmulqdq}. Our AVX-512+GFNI builds additionally use \texttt{-mgfni}. End-to-end KEM, multiplication, and component measurements are obtained from builds without internal timing code. Reed-Muller and Reed-Solomon phase breakdowns are measured from separate builds with timing code around each phase and are used only in Table~\ref{tab:decoder-cycles}.

\subsection{Component Performance}
\label{ssec:eval-component}

Cycle counts of the individual functions were presented alongside each component in Sections~\ref{sec:multiplication}--\ref{sec:keccak}. Table~\ref{tab:mul-cycles-avx512} compares AVX-512 single polynomial multiplication against Robert-V\'eron~\cite{RV23} and Cabral et al.~\cite{cabral2026accelerating}. Table~\ref{tab:decoder-cycles} compares the Reed-Muller and Reed-Solomon decoders on AVX2 and AVX-512. Table~\ref{tab:keccak-cycles} compares the SHA3-512 oracles and support expansion. Using the clean total rows, our Reed-Solomon decoder reduces the cost by 84.7\%--88.8\% over the HQC AVX2 implementation and by 88.9\%--91.0\% over Cabral et al. on AVX-512. Our Reed-Muller decoder reduces the cost by 67.9\%--77.9\% over the HQC AVX2 implementation and by 3.7\%--9.9\% over Cabral et al. on AVX-512. The AVX-512 implementations of $G$, $I$, and \texttt{vect\_write\_support\_to\_vector} reduce the cost by 37.4\%--46.4\%, 46.2\%--46.7\%, and 34.5\%--36.8\%, respectively.

\subsection{KEM Performance}
\label{ssec:eval-kem}

Table~\ref{tab:kem-cycles-avx512} reports the end-to-end cycle counts of HQC key generation, encapsulation, and decapsulation on AVX-512. We do not include SFAFFT~\cite{cryptoeprint:2026/1588} because its paper does not provide an AVX-512 implementation, and its source code was unavailable during our experiments. A comparison under the same conditions remains future work. Ours (TC/Karat) applies the optimizations of Sections~\ref{sec:multiplication}--\ref{sec:keccak} with the Toom-Cook/Karatsuba multiplication of Section~\ref{sssec:mul-tckarat}. Ours (FAFFT) uses the same implementation except that polynomial multiplication is replaced with the FAFFT implementation of Section~\ref{sssec:mul-fafft}. Ours (TC/Karat) reduces the cycle counts over Cabral et al.~\cite{cabral2026accelerating} by 24.9\%--29.5\% for key generation, 10.4\%--11.1\% for encapsulation, and 20.6\%--26.1\% for decapsulation.

At \HQC{5}, Ours (FAFFT) reduces key generation, encapsulation, and decapsulation by 5.0\%, 6.7\%, and 6.1\%, respectively, over Ours (TC/Karat). This improvement follows from the 4.1\% lower multiplication cost reported in Section~\ref{ssec:mul-comparison}. At \HQC{1} and \HQC{3}, Ours (TC/Karat) remains faster because the FAFFT transform cost is larger at these parameter sizes.

Table~\ref{tab:kem-cycles-avx2} reports the AVX2 decapsulation cycle counts against the HQC AVX2 implementation~\cite{NISTPQC-R4:HQC22}, Jang et al.~\cite{Jang26}, and Chen et al.~\cite{cryptoeprint:2026/014}. Chen et al. additionally present a cached variant that reuses the transform of $\mathbf{r}_2$ for both products and stores the transforms of $\mathbf{h}$ and $\mathbf{s}$ across calls. For comparison with Ours, the row marked $\dagger$ retains the reuse of the transform of $\mathbf{r}_2$ but does not store the transforms of $\mathbf{h}$ and $\mathbf{s}$ across calls. The unmarked ``Chen et al.'' row reports their main implementation. Ours applies the AVX2 Reed-Muller and Reed-Solomon decoding optimizations of Section~\ref{sec:decoding}. It uses the Jang et al.~\cite{Jang26} multiplication at \HQC{1} and \HQC{3}, and the AVX2 FAFFT implementation with the reduced conversion range at \HQC{5}. Ours reduces decapsulation by 11.1\%--21.5\% over the fastest prior AVX2 result for each parameter set. The reduction at \HQC{1} and \HQC{3} mainly comes from the GFNI-based Reed-Solomon decoder and the \texttt{find\_peaks} implementation of Section~\ref{sssec:rm-ours}. At \HQC{5}, the AVX2 FAFFT implementation with the reduced conversion range provides an additional reduction.

Key generation benefits from the AVX-512 oracle $I$, the use of \texttt{shake256\_512} to generate the public vector $\mathbf{h}$, and polynomial multiplication. Encapsulation benefits from polynomial multiplication, the SHA3-512 oracle $G$, and fixed-weight sampling. Decapsulation additionally benefits from the optimized Reed-Muller and Reed-Solomon decoders.

\begin{table}[!h]
\centering
\begin{threeparttable}
\caption{Cycle counts of HQC decapsulation on AVX2.}
\label{tab:kem-cycles-avx2}
\setlength{\tabcolsep}{4pt}
\begin{tabular}{l l r r r}
\toprule
Operation & Implementation & \multicolumn{1}{r}{\HQC{1}} & \multicolumn{1}{r}{\HQC{3}} & \multicolumn{1}{r}{\HQC{5}} \\
\midrule
\multirow{5}{*}{decapsulation}
  & \cite{NISTPQC-R4:HQC22} & 405{,}827 & 811{,}787 & 1{,}569{,}371 \\
  & \cite{Jang26}        & 396{,}427 & 818{,}611 & 1{,}482{,}364 \\
  & \cite{cryptoeprint:2026/014} & 403{,}807 & 831{,}652 & 1{,}278{,}585 \\
  & \cite{cryptoeprint:2026/014}$^\dagger$ & 396{,}546 & 813{,}573 & 1{,}242{,}655 \\
  & Ours           & 311{,}187 & 684{,}708 & 1{,}105{,}270 \\
\bottomrule
\end{tabular}
\begin{tablenotes}[flushleft]
\footnotesize
\item[$\dagger$] Reuses the transform of $\mathbf{r}_2$ for the two polynomial products.
\end{tablenotes}
\end{threeparttable}
\end{table}

\begin{table}[!h]
\centering
\caption{Cycle counts of HQC KEM operations on AVX-512.}
\label{tab:kem-cycles-avx512}
\setlength{\tabcolsep}{4pt}
\begin{tabular}{l l r r r}
\toprule
Operation & Implementation & \multicolumn{1}{r}{\HQC{1}} & \multicolumn{1}{r}{\HQC{3}} & \multicolumn{1}{r}{\HQC{5}} \\
\midrule
\multirow{4}{*}{key generation}
  & \cite{RV23}                   & 63{,}430 & 144{,}125 & 275{,}392 \\
  & \cite{cabral2026accelerating} & 65{,}670 & 148{,}789 & 289{,}996 \\
  & Ours (TC/Karat)         & 48{,}257 & 111{,}724 & 204{,}420 \\
  & Ours (FAFFT)            & 60{,}221 & 132{,}985 & 194{,}107 \\
\midrule
\multirow{4}{*}{encapsulation}
  & \cite{RV23}                   & 143{,}942 & 321{,}527 & 604{,}048 \\
  & \cite{cabral2026accelerating} & 101{,}345 & 229{,}154 & 441{,}300 \\
  & Ours (TC/Karat)         & 90{,}557 & 203{,}740 & 395{,}605 \\
  & Ours (FAFFT)            & 114{,}147 & 253{,}005 & 369{,}206 \\
\midrule
\multirow{4}{*}{decapsulation}
  & \cite{RV23}                   & 376{,}412 & 758{,}571 & 1{,}395{,}975 \\
  & \cite{cabral2026accelerating} & 214{,}674 & 434{,}076 & 837{,}664 \\
  & Ours (TC/Karat)         & 158{,}685 & 344{,}760 & 647{,}488 \\
  & Ours (FAFFT)            & 194{,}590 & 416{,}355 & 607{,}767 \\
\bottomrule
\end{tabular}
\end{table}

{
\section{TLS 1.3 Integration and Evaluation on x86 IoT Gateways}
\label{sec:tls}

Section~\ref{sec:evaluation} measures each KEM operation in isolation. This section evaluates the same implementation configurations in TLS 1.3. We measure handshake latency, server and client CPU costs, and the effects of network RTT and bandwidth.

\subsection{Setup}
\label{ssec:tls-setup}

We integrate each implementation into liboqs 0.16.0 and expose it to OpenSSL 3.3.2 through oqs-provider 0.12.0-dev~\cite{liboqs}. We keep each implementation's original source tree unchanged and compile it as a separate position-independent static library with the flags of Section~\ref{ssec:eval-setup}. We merge the objects into one relocatable object and link it into liboqs. Only the KEM API symbols \texttt{crypto\_kem\_keypair}, \texttt{crypto\_kem\_enc}, and \texttt{crypto\_kem\_dec} are exported. All remaining symbols are localized. Link-time optimization is disabled for every variant in this pipeline. This is the only deviation from the flags of Section~\ref{ssec:eval-setup}. For every variant and parameter set, \texttt{kat\_kem} reproduces the liboqs known-answer hashes. Thus, all stacks implement the same HQC~v5.0.0 specification.

The rows labeled ``Ours'' use the compositions of Section~\ref{ssec:eval-kem}. At levels 1 and 3, the AVX2 configuration combines the multiplication of Jang et al.~\cite{Jang26} with our decoding optimizations. At level 5, it combines the FAFFT multiplication of Chen et al.~\cite{cryptoeprint:2026/014} with the reduced conversion range and our decoding optimizations. The AVX-512 rows ``Ours (TC/Karat)'' and ``Ours (FAFFT)'' use the multiplication implementations of Section~\ref{sec:multiplication}.

We run the measurements on the Tiger Lake machine of Section~\ref{ssec:eval-setup} under Ubuntu 24.04 on WSL2. Turbo Boost is disabled, and the clock is fixed at 2.4~GHz. The server and client are pinned to distinct physical cores. The server is \texttt{openssl s\_server} with an ECDSA P-256 certificate and TLS 1.3 only. The client performs sequential full handshakes. Each iteration uses a fresh session without resumption and enables \texttt{TCP\_NODELAY}. We measure TCP connection setup and the TLS handshake separately. Unless stated otherwise, each configuration uses 10 independent runs of 200 handshakes after warmup. We report the median of the 10 run medians.

End-to-end handshake time is measured at the client from the start of \texttt{SSL\_connect}, after TCP connection setup, to its successful return. In addition, we report server and client CPU costs as on-CPU time. The server processes connections sequentially and waits for the client Finished message. Because of this wait, its wall-clock time per connection includes client computation and network delay. For server and client CPU costs, we divide the CPU time accumulated by each endpoint process by the number of handshakes. We record the server time across a run and the client time around the handshake loop, which excludes provider initialization. The client performs key generation and decapsulation, whereas the server performs encapsulation. We compare HQC with the x86-64 assembly implementation of OpenSSL X25519 and with ML-KEM through the liboqs mlkem-native backend~\cite{liboqs}, which uses AVX2 on this platform.

\subsection{Handshake Performance}
\label{ssec:tls-latency}

Table~\ref{tab:tls-loopback} reports TLS 1.3 handshake performance measured over loopback. The ``End-to-end'' columns report handshake time, while the ``Server CPU'' and ``Client CPU'' columns report CPU costs. The liboqs reference implementation of HQC yields handshakes of 18.4~ms, 50.5~ms, and 121.0~ms for the three parameter sets. Our AVX-512 implementation reaches 1.57~ms, 2.05~ms, and 2.96~ms, respectively.

Compared with Cabral et al., our AVX-512 TC/Karat implementation reduces handshake time by 4.6\% at \HQC{1}, 5.8\% at \HQC{3}, and 8.0\% at \HQC{5}. The \HQC{1} difference varied across measurement sessions, whereas the reductions at \HQC{3} and \HQC{5} were robust. Our AVX-512 FAFFT implementation provides the lowest \HQC{5} handshake time at 2.92~ms.

At \HQC{1}, our AVX-512 TC/Karat implementation completes the handshake in 1.57~ms, compared with 1.62~ms for the non-SIMD OpenSSL x86-64 assembly baseline. The AVX2 implementation of Faz-Hern\'andez et al.~\cite{FazHernandez2019ECC} and the AVX-512-IFMA implementation of ENG25519~\cite{eng25519} take 1.51~ms and 1.49~ms, respectively. Thus, our AVX-512 TC/Karat implementation is 5.1\% slower than ENG25519, the fastest classical implementation in Table~\ref{tab:tls-loopback}.

The AVX2 mlkem-native backend remains faster at matched security. The handshake times of our AVX-512 TC/Karat implementation are 12\%, 40\%, and 97\% higher at \HQC{1}, \HQC{3}, and \HQC{5}, respectively. These differences reflect the different computational costs and protocol message sizes of HQC and ML-KEM. Section~\ref{ssec:tls-netem} shows that HQC message sizes dominate handshake time on constrained links.

Reported AVX-512-optimized ML-KEM versions cannot be included because the source code of the versions reported by Zheng et al.~\cite{Zheng2024MLKEMTLSAVX512} and He et al.~\cite{pqmagic25} is unavailable. In addition, multi-way implementations such as the 32-way AVX-512 Kyber of Yu et al.~\cite{MultiWayKyber24} process many independent operations together. They are designed to improve total throughput, not the latency of a single handshake.

The ``End-to-end'' time does not show how the handshake cost divides between the two roles. The ``Server CPU'' and ``Client CPU'' columns of Table~\ref{tab:tls-loopback} show the per-role on-CPU costs. The client cost is larger than the server cost in every HQC configuration because the client performs key generation and decapsulation, whereas the server performs encapsulation. At \HQC{1}, our AVX-512 TC/Karat implementation has a server on-CPU cost of 736~$\mu$s per handshake.

\begin{table*}[!t]
\centering
\begin{threeparttable}
\caption{TLS 1.3 handshake performance over loopback in $\mu$s. End-to-end
time is the median of 10 run medians.}
\label{tab:tls-loopback}
\setlength{\tabcolsep}{4pt}
\begin{tabular}{l l l rrr rrr rrr}
\toprule
\multirow{2}{*}{ISA} & \multirow{2}{*}{Scheme} & \multirow{2}{*}{Implementation} & \multicolumn{3}{c}{End-to-end} & \multicolumn{3}{c}{Server CPU} & \multicolumn{3}{c}{Client CPU} \\
\cmidrule(lr){4-6} \cmidrule(lr){7-9} \cmidrule(lr){10-12}
 & & & \multicolumn{1}{r}{Level 1$^\S$} & \multicolumn{1}{r}{Level 3} & \multicolumn{1}{r}{Level 5} & \multicolumn{1}{r}{Level 1} & \multicolumn{1}{r}{Level 3} & \multicolumn{1}{r}{Level 5} & \multicolumn{1}{r}{Level 1} & \multicolumn{1}{r}{Level 3} & \multicolumn{1}{r}{Level 5} \\
\midrule
\multirow{2}{*}{x86$^\ddagger$} & HQC & liboqs ref.~\cite{liboqs} & 18{,}356 & 50{,}489 & 121{,}014 & 7{,}896 & 21{,}729 & 51{,}240 & 10{,}560 & 29{,}060 & 69{,}367 \\
\cmidrule(l){2-12}
     & X25519 & OpenSSL x86-64 asm & 1{,}619 & -- & -- & 732 & -- & -- & 794 & -- & -- \\
\midrule
\multirow{7}{*}{AVX2} & \multirow{5}{*}{HQC} & NIST submission~\cite{NISTPQC-R4:HQC22} & 1{,}801 & 2{,}425 & 3{,}712 & 790 & 968 & 1{,}255 & 970 & 1{,}261 & 1{,}707 \\
 & & Jang et al.~\cite{Jang26} & 1{,}787 & 2{,}418 & 3{,}564 & 793 & 959 & 1{,}211 & 968 & 1{,}245 & 1{,}650 \\
 & & Chen et al.~\cite{cryptoeprint:2026/014} & 1{,}864 & 2{,}483 & 3{,}421 & 823 & 1{,}017 & 1{,}179 & 999 & 1{,}284 & 1{,}547 \\
 & & Chen et al.$^\dagger$~\cite{cryptoeprint:2026/014} & 1{,}858 & 2{,}454 & 3{,}381 & 825 & 993 & 1{,}167 & 1{,}007 & 1{,}267 & 1{,}540 \\
 & & Ours & 1{,}737 & 2{,}332 & 3{,}307 & 787 & 958 & 1{,}145 & 924 & 1{,}174 & 1{,}456 \\
\cmidrule(l){2-12}
 & ML-KEM & mlkem-native~\cite{liboqs} & 1{,}399 & 1{,}462 & 1{,}502 & 654 & 663 & 676 & 709 & 728 & 744 \\
\cmidrule(l){2-12}
 & X25519 & Faz-Hern\'andez et al.~\cite{FazHernandez2019ECC} & 1{,}514 & -- & -- & 695 & -- & -- & 738 & -- & -- \\
\midrule
\multirow{5}{*}{AVX-512} & \multirow{4}{*}{HQC} & Robert-V\'eron~\cite{RV23} & 1{,}709 & 2{,}313 & 3{,}370 & 778 & 904 & 1{,}135 & 958 & 1{,}187 & 1{,}563 \\
 & & Cabral et al.~\cite{cabral2026accelerating} & 1{,}645 & 2{,}178 & 3{,}222 & 750 & 898 & 1{,}088 & 891 & 1{,}106 & 1{,}389 \\
 & & Ours (TC/Karat) & 1{,}570 & 2{,}052 & 2{,}963 & 736 & 875 & 1{,}061 & 853 & 1{,}029 & 1{,}282 \\
 & & Ours (FAFFT) & 1{,}652 & 2{,}165 & 2{,}923 & 791 & 904 & 1{,}016 & 918 & 1{,}078 & 1{,}236 \\
\cmidrule(l){2-12}
 & X25519 & ENG25519~\cite{eng25519} & 1{,}494 & -- & -- & 686 & -- & -- & 744 & -- & -- \\
\bottomrule
\end{tabular}
\begin{tablenotes}[flushleft]
\footnotesize
\item[$\S$] Columns give the NIST security level, namely \HQC{1}, \HQC{3},
and \HQC{5} for HQC and ML-KEM-512, ML-KEM-768, and ML-KEM-1024 for ML-KEM.
X25519 provides 128-bit classical security and is listed at level 1 for
reference.
\item[$\ddagger$] Base x86 instruction set without SIMD extensions.
\item[$\dagger$] Reuses the transform of $\mathbf{r}_2$ for the two polynomial products.
\end{tablenotes}
\end{threeparttable}
\end{table*}

\subsection{Network Performance}
\label{ssec:tls-netem}

IoT networks are often constrained by both latency and bandwidth. We therefore evaluate paths with added RTTs of 0 and 50~ms using \texttt{netem}, with no rate limit or rate limits of 10~Mbit/s and 1~Mbit/s. The emulated path uses a virtual Ethernet pair with an MTU of 1500 bytes. Table~\ref{tab:tls-bandwidth} reports median handshake times over 100 handshakes.

Without a rate limit, the local computation differences remain visible at added RTT 0. With no rate limit and an added RTT of 50~ms, \HQC{1} and \HQC{3} take about 52~ms in every optimized configuration, whereas \HQC{5} takes about 104~ms. The 14{,}421-byte \HQC{5} ciphertext increases the server flight to about 15.5~KB, which exceeds the default TCP initial congestion window of 10 segments at this MTU. Thus, the large \HQC{5} ciphertext can cause an additional round trip by exceeding the initial congestion window, as observed on the faster paths. Sosnowski et al.~\cite{Sosnowski23} report the same congestion-window effect for post-quantum handshakes with large messages.

At 10~Mbit/s and 1~Mbit/s, the time to transmit the handshake messages increasingly dominates. At 1~Mbit/s, the slower transmission lets acknowledgments arrive before the initial window is exhausted, eliminating the distinct RTT-scale penalty seen on faster paths. Under this rate limit, all HQC implementations take about 193~ms at added RTT 0 and 243~ms at added RTT 50 for \HQC{5}, while ML-KEM-1024 takes 37~ms and 87~ms, respectively. Thus, on constrained links, the communication sizes of HQC, rather than the choice between the AVX2 and AVX-512 implementations, determine the overall handshake time.

For Table~\ref{tab:tls-bandwidth}, ``Ours (AVX-512)'' uses TC/Karat at all levels. ``Ours (AVX2)'' uses TC/Karat at \HQC{1} and \HQC{3}, and FAFFT at \HQC{5}.

\begin{table*}[!t]
\centering
\begin{threeparttable}
\caption{Median TLS 1.3 handshake time in ms over emulated paths with added RTT and bandwidth constraints. Each entry is the median of 100 handshakes.}
\label{tab:tls-bandwidth}
\scriptsize
\setlength{\tabcolsep}{2pt}
\begin{tabular}{c l rrr rrr rrr rrr r}
\toprule
& & \multicolumn{3}{c}{\shortstack{Cabral et al.~\cite{cabral2026accelerating}\\(AVX-512)}} & \multicolumn{3}{c}{\shortstack{Ours\\(AVX-512)}} & \multicolumn{3}{c}{\shortstack{Ours\\(AVX2)}} & \multicolumn{3}{c}{\shortstack{ML-KEM\\(AVX2)}} & \\
\cmidrule(lr){3-5} \cmidrule(lr){6-8} \cmidrule(lr){9-11} \cmidrule(lr){12-14}
RTT & Rate & \HQC{1} & \HQC{3} & \HQC{5} & \HQC{1} & \HQC{3} & \HQC{5} & \HQC{1} & \HQC{3} & \HQC{5} & -512 & -768 & -1024 & X25519 \\
\midrule
\multirow{3}{*}{0~ms} & No rate limit & 1.14 & 1.37 & 1.76 & 1.09 & 1.08 & 1.67 & 1.18 & 1.52 & 1.90 & 1.11 & 1.13 & 1.16 & 1.23 \\
& 10~Mbit/s & 7.91 & 13.83 & 21.43 & 7.89 & 13.93 & 21.43 & 7.87 & 13.94 & 21.98 & 3.74 & 4.40 & 5.27 & 2.50 \\
& 1~Mbit/s & 66.89 & 124.37 & 193.34 & 66.83 & 124.15 & 193.20 & 66.96 & 124.37 & 193.47 & 23.86 & 29.61 & 36.94 & 11.08 \\
\midrule
\multirow{3}{*}{50~ms} & No rate limit & 52.39 & 52.92 & 104.37 & 52.23 & 52.05 & 104.16 & 52.50 & 52.65 & 104.66 & 52.29 & 52.35 & 52.40 & 52.54 \\
& 10~Mbit/s & 58.58 & 64.37 & 113.42 & 58.42 & 64.43 & 113.32 & 58.43 & 64.70 & 113.64 & 54.60 & 55.22 & 56.08 & 53.33 \\
& 1~Mbit/s & 116.89 & 174.07 & 243.41 & 116.77 & 174.23 & 243.04 & 116.90 & 174.23 & 243.52 & 74.08 & 79.82 & 87.44 & 61.63 \\
\bottomrule
\end{tabular}
\end{threeparttable}
\end{table*}

\subsection{Storage and Memory Cost}
\label{ssec:tls-footprint}

The liboqs shared library built from the reference implementation occupies 0.77~MB. Our AVX-512 TC/Karat build occupies 1.14~MB, close to the 1.04~MB of Cabral et al., while every FAFFT-based build occupies 2.5~MB--2.75~MB due to transform tables. The performance results favor TC/Karat at \HQC{1} and \HQC{3}, and FAFFT at \HQC{5}. For an \HQC{5} deployment with limited storage space, TC/Karat remains a smaller alternative to FAFFT. Peak resident memory of the TLS endpoints is 9 to 10~MB in every configuration and is dominated by OpenSSL itself, so the choice of HQC implementation does not change process memory in practice.

}
\section{Conclusion}
\label{sec:conclusion}

This paper presented optimized HQC implementations for AVX2 and AVX-512 + GFNI, and evaluated them in TLS 1.3 on x86 IoT gateways. In these implementations, we extended GFmulOpt to AVX-512 for polynomial multiplication and used the AVX-512 base multiplication of Robert and V\'eron~\cite{RV23} inside the selected decomposition trees. We also ported FAFFT multiplication to AVX-512 and reduced its transform cost with precomputed twiddle tables and reduced basis conversions. For decoding, we optimized both Reed-Muller and Reed-Solomon decoding in the AVX2 and AVX-512 settings by combining vector data movement with GFNI-based arithmetic over $\mathbb{F}_{2^8}^{\mathrm{RS}}$. For the Keccak-based functions, we implemented the SHA3-512 oracles with AVX-512 and ported support expansion in fixed-weight sampling to AVX-512.

At the KEM level, our AVX-512 TC/Karat implementation reduces key generation, encapsulation, and decapsulation by 24.9\%--29.5\%, 10.4\%--11.1\%, and 20.6\%--26.1\%, respectively, relative to Cabral et al. Our AVX2 implementation reduces decapsulation by 11.1\%--21.5\% relative to the fastest prior AVX2 result.

For TLS 1.3, we measure end-to-end handshake latency and client and server on-CPU cost on loopback, then evaluate paths with added RTT and bandwidth constraints. On loopback, our AVX-512 TC/Karat implementation reduces the median handshake time from 2.18~ms to 2.05~ms at \HQC{3} relative to Cabral et al., while our AVX-512 FAFFT implementation reduces it from 3.22~ms to 2.92~ms at \HQC{5}. With a 50~ms added RTT and no rate limit, \HQC{1} and \HQC{3} take about 52~ms, whereas \HQC{5} takes about 104~ms. At 1~Mbit/s, \HQC{5} takes about 193~ms and 243~ms at added RTTs of 0 and 50~ms, compared with 37~ms and 87~ms for ML-KEM-1024. Our measurements confirm that, on constrained links, the communication sizes of HQC determine the overall handshake time more than the HQC implementation.

Our evaluation has two limitations. We use a single Tiger Lake processor with Turbo Boost and Hyper-Threading disabled, and the TLS tests target x86 IoT gateways over emulated network paths. Future work includes evaluating the implementations on additional server platforms, resource-constrained end devices, and deployed wireless IoT networks. For the HQC implementation, future work includes applying SFAFFT~\cite{cryptoeprint:2026/1588} to AVX-512 and integrating the caching techniques studied by Chen et al.~\cite{cryptoeprint:2026/014} and Jang et al.~\cite{Jang26} to evaluate the tradeoff between memory use and cycle counts.

\appendix[Note on the HQC Update of October 2026]
\label{app:spec-revision}

The implementations and measurements in this paper follow the HQC specification of August 22, 2025, denoted v5.0.0, and the known-answer tests distributed with it. After our experiments were completed, the HQC team published a revised specification on October 12, 2026~\cite{HQCspec2026}. The reference implementation and the known-answer tests of the revision were not available at the time of writing. This appendix summarizes the revision based on the specification document and describes its effect on the results of this paper.

The revision changes three items that affect implementations. First, the weight $\omega$ increases by one for all three parameter sets, and $\omega_r=\omega_e$ increases by one for \HQC{1} and \HQC{5}. The length $N$, the code parameters $n_1$, $n_2$, and $\delta$, and the sizes of the encapsulation key and the ciphertext are unchanged. Second, the Reed-Solomon field is redefined with the irreducible polynomial $x^8+x^4+x^3+x+1$ and therefore coincides with $\mathbb{F}_{2^8}^{\mathrm{AES}}$ of Section~\ref{ssec:notation}. The generator polynomials listed in the revision correspond to the primitive element $x+1$, and the Reed-Solomon generator polynomials change accordingly. Third, the default decapsulation key is reduced to the 32-byte seed $\mathit{seed}_{\mathsf{KEM}}$. Decapsulation first recomputes $(\mathsf{ek}_{\mathsf{PKE}},\mathsf{dk}_{\mathsf{PKE}})$ and $\sigma$ from this seed by running \textsf{HQC-PKE.KeyGen} and the XOF of Algorithm~\ref{alg:hqc-kem}, and then proceeds as before. The sampling algorithms, the sampling order, and the oracles $G$, $H$, $I$, and $J$ are unchanged.

Polynomial multiplication in Section~\ref{sec:multiplication}, the Reed-Muller decoder in Section~\ref{ssec:rm}, and the Keccak-based functions in Section~\ref{sec:keccak} do not depend on the changed items and apply without modification. Support expansion in Section~\ref{sssec:keccak-sampling} scans all output words against the support positions, so its cost grows by one position per sampled vector.

The Reed-Solomon decoder in Section~\ref{sssec:rs-ours} keeps its structure, and the precomputed tables, namely the generator polynomial, the syndrome evaluation points, and the Chien search candidates, are regenerated for the revised field. Since the revised field coincides with $\mathbb{F}_{2^8}^{\mathrm{AES}}$, the \texttt{vgf2p8affineqb} conversions of Section~\ref{ssec:avx512-gfni} before and after each GFNI operation become unnecessary. A packed multiplication then requires a single \texttt{vgf2p8mulb} instead of \texttt{vgf2p8mulb} with surrounding conversions, and an inversion requires a single \texttt{vgf2p8affineinvqb}. The decoder cost under the revision is therefore bounded by the values in Table~\ref{tab:decoder-cycles}. The same simplification applies to the syndrome computation of Chen et al.~\cite{cryptoeprint:2026/014}, whereas the \texttt{PCLMULQDQ}-based field arithmetic of the HQC AVX2 implementation~\cite{NISTPQC-R4:HQC22} and Cabral et al.~\cite{cabral2026accelerating} changes only its reduction constants.

Under the compressed decapsulation key, decapsulation repeats key generation except for the sampling of $\mathit{seed}_{\mathsf{KEM}}$, namely the XOF expansion of the seed, the oracle $I$, the sampling of $\mathbf{y}$ and $\mathbf{x}$, the expansion of $\mathbf{h}$, and the product $\mathbf{h}\cdot\mathbf{y}$. The sum of the key generation and decapsulation rows of Table~\ref{tab:kem-cycles-avx512} therefore approximates the decapsulation cost of the revised algorithm. The two differ only by the sampling of $\mathit{seed}_{\mathsf{KEM}}$ and by the serialization and parsing of the original decapsulation key. For Ours (TC/Karat), this sum is $206{,}942$, $456{,}484$, and $851{,}908$ cycles at \HQC{1}, \HQC{3}, and \HQC{5}, compared with $280{,}344$, $582{,}865$, and $1{,}127{,}660$ cycles for Cabral et al. At \HQC{5}, the sum for Ours (FAFFT) is $801{,}874$ cycles. The revision notes that implementations may cache the expanded key to avoid this overhead.

In TLS 1.3, the client performs key generation and decapsulation in each handshake, as described in Section~\ref{ssec:pqtls}. Under the compressed decapsulation key, the client CPU cost in Table~\ref{tab:tls-loopback} increases by about one key generation unless the integration retains the expanded key between the two operations, which the revision permits. The server cost and the sizes of the ClientHello and the ciphertext are unchanged, so the RTT- and bandwidth-dependent results of Section~\ref{ssec:tls-netem} are unaffected.

\bibliographystyle{IEEEtran}
\bibliography{ref}

\end{document}